\documentclass[11pt,twocolumn]{emulateapj_Astroph}

\tightenlines
\usepackage{epstopdf}
\usepackage{ulem} % AV
\usepackage{color} % AV

\def\lsim{\;\raise0.3ex\hbox{$<$\kern-0.75em\raise-1.1ex\hbox{$\sim$}}\;}
\def\gsim{\;\raise0.3ex\hbox{$>$\kern-0.75em\raise-1.1ex\hbox{$\sim$}}\;}

\definecolor{purple}{RGB}{200,100,255} %{255,100,20}

\newcommand{\aafourUM}{UM}
\newcommand{\aafour}{A}
\newcommand{\aafive}{B}
\newcommand{\aathree}{C}
\newcommand{\aatwo}{D}
\newcommand{\aaeleven}{E}
\newcommand{\aaseven}{F}
\newcommand{\aaone}{G}
\newcommand{\aasixteen}{H}
\newcommand{\aasix}{I}
\newcommand{\aaeight}{J}
\newcommand{\aanine}{K}
\newcommand{\aaten}{L}
\newcommand{\aafourteen}{M}
\newcommand{\bbthree}{N}
\newcommand{\aaseventeen}{P}
\newcommand{\bbfour}{Q}
\newcommand{\bbeight}{R}
\newcommand{\bbsix}{S}
\newcommand{\bbtwo}{T}
\newcommand{\bbten}{U}
\newcommand{\bbtwelve}{V}
\newcommand{\bbeleven}{W}
\newcommand{\bbseven}{X}
\newcommand{\bbfive}{Y}
\newcommand{\bbone}{Z}
\newcommand{\bbnine}{AA}
\newcommand{\degg}{^\circ}

\newcommand{\Kol}{Kolmogorov}
\newcommand{\Lmax}{L_\mathrm{max}}
\newcommand{\Tmax}{t_\mathrm{max}}
\newcommand{\Btot}{B_\mathrm{tot}}

\newcommand{\Opi}{\omega_\mathrm{pi}}

\newcommand{\pTran}{p_\mathrm{tran}}
\newcommand{\fTran}{f_\mathrm{tran}}
\newcommand{\sas}{small-angle-scattering}

\newcommand{\RT}{Raleigh--Taylor}
\newcommand{\nonres}{non-resonant}

\newcommand{\Valf}{v_{a}}
\newcommand{\xx}[1]{\!\times\!10^{#1}}
\newcommand{\Rsk}{R_\mathrm{sk}}
\newcommand{\Lfeb}{L_\mathrm{FEB}}
\newcommand{\pcc}{cm$^{-3}$}
\newcommand{\kmps}{km s$^{-1}$}

\newcommand{\BB}{broadband}
\newcommand{\muG}{$\mu$G}
\newcommand{\DSAshock}{DSA/$\Delta B$/smooth--shock}
\newcommand{\VFP}{Vlasov--Fokker--Planck}
\newcommand{\SA}{semi-analytic}
\newcommand{\FEB}{free escape boundary}

\newcommand{\nCR}{n_\mathrm{cr}}
\newcommand{\RH}{Rankine-Hugoniot}
\newcommand{\rRH}{R_\mathrm{RH}}

\newcommand{\Qlin}{quasi-linear}
\newcommand{\mfp}{mean free path}
\newcommand{\Bss}{B_\mathrm{ss}}
\newcommand{\Kss}{k_\mathrm{ss}}
\newcommand{\Rss}{r_\mathrm{ss}}

\newcommand{\mfpTH}{\lambda_\mathrm{th}}
\newcommand{\Lpic}{\lambda_\mathrm{pic}}
\newcommand{\Lss}{\lambda_\mathrm{ss}}
\newcommand{\Kvor}{k_\mathrm{vor}}
\newcommand{\LamVor}{\lambda_\mathrm{vor}}
\newcommand{\Lvor}{l_\mathrm{vor}}
\newcommand{\Dvor}{D_\mathrm{vor}}
\newcommand{\Uvor}{U_\mathrm{vor}}

\newcommand{\Bls}{B_\mathrm{ls}}

\newcommand{\Bkolm}{B_\mathrm{Kolm}}
\newcommand{\Bism}{B_\mathrm{ISM}}

\newcommand{\Beff}{B_\mathrm{eff}}
\newcommand{\BeffDS}{B_\mathrm{eff,2}}
\newcommand{\Lls}{L_\mathrm{ls}}
\newcommand{\Kls}{k_\mathrm{ls}}
\newcommand{\Lcor}{l_\mathrm{cor}}
\newcommand{\Lres}{\lambda_\mathrm{res}}
\newcommand{\Kres}{k_\mathrm{res}}
\newcommand{\thetaScat}{\theta_\mathrm{scat}}
\newcommand{\rg}{r_g}
\newcommand{\rgz}{r_{g0}}
\newcommand{\rgZ}{r_{g0}}
\newcommand{\rgTH}{r_{g,\mathrm{th}}}

\newcommand{\deltime}{\delta t}
\newcommand{\Lmfp}{\lambda_\mathrm{mfp}}
\newcommand{\Rtot}{R_\mathrm{tot}}
\newcommand{\Rsub}{R_\mathrm{sub}}
\newcommand{\VscatI}{v_\mathrm{scat}^{(i+1)}}
\newcommand{\Vscat}{v_\mathrm{scat}}

\newcommand{\Jcr}{J_\mathrm{cr}}
\newcommand{\JcrI}{J^{\mathrm{cr}}}
\newcommand{\fcr}{f_\mathrm{cr}}
\newcommand{\Fth}{F_\mathrm{th}}
\newcommand{\Fcr}{F_\mathrm{cr}}
\newcommand{\FwI}{F_w^{(i)}}
\newcommand{\Fw}{F_w}
\newcommand{\Fwc}{{\cal F}_w}
\newcommand{\Lwc}{{\cal L}}
\newcommand{\pth}{p_\mathrm{th}}
\newcommand{\Pth}{P_\mathrm{th}}
\newcommand{\Pcr}{P_\mathrm{cr}}
\newcommand{\PcrDS}{P_\mathrm{cr,2}}
\newcommand{\Pw}{P_w}
\newcommand{\Pwc}{{\cal P}_w}
\newcommand{\PwDS}{P_{w,2}}
\newcommand{\PwI}{P_w^{(i)}}
\newcommand{\PhiEmc}{\Phi^\mathrm{({\it i})MC}_E}
\newcommand{\PhiPmc}{\Phi^\mathrm{({\it i})MC}_P}
\newcommand{\PhiP}{\Phi^\mathrm{part}_P}
\newcommand{\PhiPz}{\Phi_{P0}}
\newcommand{\PhiE}{\Phi^\mathrm{part}_E}
\newcommand{\PhiEz}{\Phi_{E0}}
\newcommand{\Qesc}{Q_\mathrm{esc}}

\newcommand{\FpxZ}{\PhiPz}
\newcommand{\nonrel}{non-relativistic}
\newcommand{\rel}{relativistic}

\newcommand{\pmax}{p_\mathrm{max}}
\newcommand{\pmin}{p_\mathrm{min}}
\newcommand{\kmax}{k_\mathrm{max}}
\newcommand{\kmin}{k_\mathrm{min}}

\newcommand{\mc}{Monte Carlo}

\newcommand{\DSA}{diffusive shock acceleration}

\newcommand{\MFA}{magnetic field amplification}
\newcommand{\alf}{Alfv\'en}
\newcommand{\Alf}{Alfv\'en}

\newcommand{\SAlf}{super-alfv\'enic}
\newcommand{\subAlf}{sub-alfv\'enic}

\newcommand{\NL}{nonlinear}
\newcommand{\SCly}{self-consistently}
\newcommand{\SC}{self-consistent}

\renewcommand{\baselinestretch}{1}
\def\aap{Astronomy and Astrophysics}

\def\apjl{The Astrophysical Journal Letters}

\newcount\listnorom
\newcommand\listromanDE{\global\advance \listnorom by 1
{\lowercase\expandafter{(\romannumeral\listnorom)}\ }}
\newcommand\newlistroman{\listnorom=0}

\newcount\listnumber
\newcommand\listDE{\global\advance \listnumber by 1
{\lowercase\expandafter{(\number\listnumber)}\ }}
\newcommand\newlistDE{\listnumber=0}
\newcount\Inum
\newcount\IInum
\newcount\IIInum
\newcount\IVnum
\def\I{\global\multiply\IInum by 0 \global\multiply\IIInum by 0
            \global\multiply\IVnum by 0 \global\advance \Inum by 1
            {\the\Inum. }}
\def\II{\global\multiply\IIInum by 0\global\multiply\IVnum by 0
       \global\advance \IInum by 1 {\the\Inum.\the\IInum. }}
\def\III{\global\multiply\IVnum by 0\global\advance \IIInum by 1
            {\the\Inum.\the\IInum.\the\IIInum. }}
\def\IV{\global\advance \IVnum by 1
            {\the\IVnum. }}
\shorttitle{Three Instability Modes and Cascading in DSA}
\shortauthors{Bykov, Ellison, Osipov, Vladimirov}

\begin{document}
%\medskip

\title{Magnetic field amplification in nonlinear diffusive shock acceleration including resonant and
non-resonant cosmic-ray driven instabilities}

\vskip24pt

\author{Andrei M. Bykov}\affil{A.F.Ioffe Physical-Technical Institute, St. Petersburg 194021, also St.Petersburg State
Politechnical University, Russia, and the International Space Science Institute, Bern, Switzerland; ambykov@yahoo.com}
\author{Donald C. Ellison}
  \affil{North Carolina State University, Department of Physics, Raleigh, NC 27695-8202, USA; don\_ellison@ncsu.edu}
\author{Sergei M. Osipov}
  \affil{A.F.Ioffe Physical-Technical Institute, St. Petersburg, Russia; osm2004@mail.ru}
\author{Andrey E. Vladimirov}
  \affil{Colfax International, 750 Palomar Ave, Sunnyvale, CA 94085, USA; avenovo@gmail.com}

%PACS numbers: 94.20.Wc, 98.38.Mz, 98.70.Sa

\begin{abstract}
We present a \NL\ \mc\ model of efficient \DSA\ (DSA) where the magnetic turbulence responsible for particle diffusion is calculated \SCly\ from the resonant cosmic-ray (CR) streaming instability, together with  \nonres\ short-- and
long--wavelength CR--current--driven instabilities.
We include the backpressure from CRs interacting with the strongly amplified magnetic turbulence which decelerates and heats the \SAlf\ flow in the extended shock precursor.
Uniquely, in our plane-parallel, steady-state, multi-scale model, the full range of particles, from thermal ($\sim$ eV) injected at the viscous subshock, to the escape of the  highest energy CRs ($\sim$ PeV) from the shock precursor, are calculated consistently with the shock structure, precursor heating, \MFA\ (MFA), and scattering center drift relative to the background plasma.
In addition, we show how  the cascade of turbulence to shorter wavelengths
influences  the total shock compression, the downstream proton temperature, the magnetic fluctuation spectra, and accelerated particle spectra.
A parameter survey is included where we vary shock parameters, the mode of  magnetic turbulence generation, and turbulence cascading.
From our survey results,
we obtain scaling relations for the maximum particle momentum and amplified magnetic field as functions of shock speed, ambient density, and shock size.
\end{abstract}

\section{Introduction}
The existence of strong, \SAlf\ collisionless shocks in many astrophysical objects, such as supernova remnants (SNRs), extra-galactic radio jets, clusters of galaxies, and compact accreting sources,
has been revealed to us by nonthermal radiation produced by \rel\ particles.
Diffusive shock acceleration (DSA), the most favored acceleration mechanism for producing these \rel\ particles
\citep[e.g.,][]{JE91,md01}, is expected to be efficient and to
produce highly non-equilibrium particle
populations.
A high acceleration efficiency implies
strong
coupling between the accelerated particle population, the shock structure, and the
electromagnetic fluctuations, with a wide range of scales,
responsible for scattering the
particles, making DSA a difficult nonlinear (NL)
problem.

Efficient DSA produces a hard energy spectrum and the
backpressure of these high-energy CRs
penetrating into the cold incoming plasma creates a  precursor with
a length-scale on the order of the diffusion length of the highest
energy CRs \citep[e.g.,][]{be87}.
Imbeded in this precursor is a short-scale, viscous subshock that is largely responsible
for heating the plasma and injecting particles into the Fermi mechanism.
Basic considerations of momentum and energy conservation
require that the production of \rel\ particles involving the largest magnetic
turbulence scales
must impact the injection of thermal particles and the
structure of the shock on the smallest ion inertial scales making the shock intrinsically multi-scale.

Because hard spectra can be produced in efficient DSA,
CRs with the longest diffusion lengths can escape at an upstream boundary and carry away
a sizable fraction of the shock ram pressure
\citep[e.g.,][]{EJE1981,berkrym88,be99,
cab10,eb11,DruryEscape11,Malkov_etal_escape12}.\footnote{Of course,
escape can occur at any boundary but for concreteness we only consider upstream escape from the shock precursor.}
This escaping energy flux plays an important
role in DSA and must be \SCly\ included in determining the shock
structure. It will cause the overall shock compression ratio to increase and the escaping CR current is certain to generate
turbulence that will serve as seed turbulence for compression and amplification as it is overtaken by the
shock \citep[e.g.,][]{BF2007,BAC2007,BellEtal2013}.

Global conservation considerations aside, the detailed formation and structure of collisionless shocks is
by no means certain.
Early on it was suggested that a Weibel-type instability from
counter streaming
plasma flows could result in the formation of a gas subshock; a narrow
region filled with strong magnetic fluctuations deflecting the
incoming particles \citep[e.g.,][]{ms63,tk71}.
This basic picture
was supported later by direct spacecraft
observations of heliospheric shocks
\citep[e.g.,][]{kennelea84,ts85,goslingea89}.

In addition to observations,
hybrid and particle-in-cell (PIC) plasma
simulations\footnote{In PIC simulations, both electrons and ions are followed while in hybrid simulations the ions are followed but the electrons are treated as a charge neutralizing fluid. Since hybrid simulations do not follow the plasma on electron time scales the computational requirements are much less.}
have investigated collisionless shocks coupled with particle acceleration.
To our knowledge, the first plasma simulation to show clear evidence of DSA was the parallel-shock, hybrid
simulation of \citet{Quest1988}.
Since then, a great deal of work with both hybrid and
full-particle PIC simulations has been done
\citep[see][ and references therein, for a sampling of this large body of work]{wq88,winskeea90,EGBS93,
giacaloneea97, kt08,kt10,spitkovsky08a,spitkovsky08b,treumann09,
gs12,bs13,CS2013}.
The direct simulation of the microscopic structure of the shock
on scales of a few thousand ion
inertial lengths\footnote{The ion inertial length is $c/\Opi$, where $\Opi=\sqrt{4 \pi e^2 n_e/m_p}$ is the proton plasma frequency, $c$ is the speed of light, $e$ is the electronic charge,  $n_e$ is the electron number density, and $m_p$ is the proton mass.}
has been
extremely important for resolving the microscopic structure of the plasma shock transition and
for clearly demonstrating the initialization of the Fermi
acceleration process.
However, PIC/hybrid simulations are computationally demanding and
the runtime, box size, number of particles, and particle momentum
range that can be simulated are limited. In \nonrel\ shock simulations,
they are currently limited to two or three decades in the dynamical
ranges of the spectra of energetic particles and in the $k$-space of
the magnetic fluctuations.

In contrast, models of strong MFA applicable to SNRs require
dynamical ranges for particle momentum and turbulence extending
nine or ten decades. Therefore, to capture the \NL\ physics
connecting the highest energy CRs to the self-generated,  \BB\ wave
turbulence, and to the viscous subshock structure, coarse grained,
multi-scale models  of the collisionless plasma turbulence
generation must be used in addition to the microscopic treatment
afforded by PIC/hybrid simulations \citep[see,
e.g.,][]{be87,md01,ab06,vbe08,kang13,rb13}.

Different approaches to model the multi-scale nature of DSA in
strong non-relativistic shocks are currently underway. Kinetic \SA\
models \citep[see,
e.g.,][]{md01,vbk05,ab06,za10,bgo13,LSENP2013,SB13,rb13} perform \SC\
calculations once the injection rate of background  particles into
the Fermi mechanism is parameterized. These models use various
approximations for the particle diffusion coefficient and the MFA.
The background plasma is typically modeled with a
diffusion-convection equation which incorporates the magnetic field
structure.
Some recent time-dependent models have been
presented by \citet{KRJ2012}, \citet{zp12}, and \citet{SB13}.

Another approach combines standard MHD equations to describe  the
background plasma, including  the CR current
$\mathbf{j_\mathrm{CR}}\times \mathbf{B}$ term producing the
non-resonant hybrid (NRH) (i.e., Bell) instability, with a \VFP\
equation describing the CR distribution function \citep*[see][ and
references therein]{BellEtal2013}.

Here, we use a \mc\ technique which allows broadband modeling of \NL\ DSA incorporating both resonant and
non-resonant (short-- and long--wavelength) current-driven
instabilities.
This is a plane-parallel, steady-state model extending
earlier work by, for example, \citet{JE91,EBJ96,veb06,vbe08,vbe09}.
The model gives  a consistent  account of the amplification of
mesoscopic scale fluctuating magnetic fields interacting with
superthermal CRs and the
NL feedback of particles and fields on the bulk flow is
derived from basic  energy-momentum conservation
laws.
By mesoscale, we mean magnetic turbulence with scales between  the
very short-scale Weibel-type instabilities that determine the
subshock structure and are seen in PIC simulations, and the very
large-scale turbulence, e.g., Raleigh--Taylor instabilities, seen
in hydro or MHD simulations
\citep[e.g.,][]{GiacJok07,FDS2012,WB2013}. Weibel and \RT\
instabilities are not included in our model.

In contrast to \SA\ models based on the diffusion-advection
equation, which make the diffusion approximation and require an
independent injection parameter, thermal particle injection in the
\mc\ model  is achieved  \SCly\ once the particle mean free path is
specified (see \S\ref{sec:mfp}).
Particles from thermal energies to energies sufficient to capture
the  essential NL effects expected from CR production in young SNRs,
as well as the magnetic turbulence interacting with these particles,
are included \SCly.

The source of the mesoscopic resonant and non-resonant turbulence in
DSA is the anisotropy of the distribution function of the energetic
particles. Since the \mc\ method does not assume near isotropic
distributions,   no approximation is made concerning the CR
distribution.
The  CR pressure gradient and current are obtained to all orders
in the shock precursor. Due to computational limits, more restrictive approximations may be necessary in models based on the Vlasov-Fokker-Planck equation \citep[e.g.,][]{BellEtal2013}.

We calculate the spectra of the magnetic turbulence using the
quasi-linear growth rates described in \S\ref{sec:Grow}. The growth
rates  are derived from a general dispersion relation given by
\citet[][]{bbmo13}. This single dispersion relation accounts for the three main CR-driven
instabilities: Bell's short-scale instability
\citep[][]{bell04,bell05},
the resonant streaming instability
\citep[see, for example,][]{achterberg81,be87},
and the long-wavelength, ponderomotive
instability \citep[see][]{boe11,bbmo13,SchureEtal2012}.

The \mc\ simulation solves the full \DSAshock\  problem iteratively. Within this process,
we calculate
the local growth rates at any iteration using the mean magnetic field and the
CR-current derived at the previous iteration.
Importantly, between any two iterations $\Delta B/B < 1$ so
we converge to an amplified magnetic field $\Delta B(x) \gg B_0$, i.e., with
rms-amplitudes much larger than the initial magnetic field $B_0$,
without violating the \Qlin\
approximation (see \S~\ref{sec:MC}).\footnote{To go from $B_0=3$\,\muG\ to $B > 300$\,\muG\
with $\Delta B/B \lsim 0.1$ requires $\sim 50$ iterations, a number
easily accommodated within  the \mc\ procedure.}

The spectrum of the mesoscopic magnetic fluctuations depends
critically on cascading, i.e., the transfer of turbulent energy from
long to short wavelengths \citep[e.g.,][]{vbe08}. If MFA is
strong, and local turbulent cascading parallel to the mean field is
suppressed, as expected in MHD turbulence,  then the  turbulence
spectrum will contain one or more discrete peaks \citep[see][and
Section~\ref{sec:MC} below]{vbe09,Tycho_stripes11}.
We show, however, that critical aspects of DSA, such as the particle distribution function, are relatively insensitive to cascading. This effect is discussed in detail in \S\ref{sec:Flux}.

In what follows we
emphasize the complex, coupled
nature of the problem necessitated by  the efficient overall
acceleration, the wide dynamic range of the CR distribution
function, and the equally \BB, self-generated
turbulence produced simultaneously by several instabilities.
To highlight these effects, we consider acceleration efficiencies where more than 50\% of the shock ram kinetic energy is placed in accelerated particles; combined in trapped and escaping CRs.
While trapped and escaping CRs are produced together from the shock ram pressure dissipation they have different properties. The escaping CRs contribute immediately to the galactic CR population while the fate of the trapped CRs depends on the shock evolution, which we do not address in our steady-state, plane geometry model.
The trapped particles will eventually leave the SNR but only after losing some fraction of their energy to work done on the magnetic field and expanding plasma.
Our model is designed to understand the physical effects of efficient DSA in the local vicinity of an extended supernova shock.

Of particular importance is our treatment of the scattering center speed, $\Vscat(x)$,  relative to the bulk
plasma speed, $u(x)$. This is
determined from basic conservation considerations without assuming \alf\ waves and we find $|\Vscat|$ to be significantly below what \Alf\ waves would predict.
Finally, a limited parameter survey is given and we discuss the parameters that determine the maximum CR energy a given shock can produce as well as various observational consequences related to the magnetic turbulence spectra we calculate.

\section{Model}
The basic elements of our plane-parallel, steady-state,
\mc\ method have been discussed in a number of previous papers
\citep[e.g.,][]{JE91,EBJ96,EWB2013}. Support for its  general validity comes from
detailed comparisons
with {\it in situ} spacecraft observations of particle acceleration at heliospheric shocks
\citep[e.g.,][]{be87,EMP90,BOEF97}, as well as from direct comparison with
hybrid  plasma simulations \citep[][]{EGBS93}.
While early versions of the \mc\ code asummed simple forms for the
particle  diffusion, we continue here an extensive generalization of
the code to include magnetic turbulence generation and \SC\
diffusion \citep[e.g.,][]{veb06,vbe08,vbe09}.

The \mc\ code includes the following main elements: \\
$\bullet$ thermal leakage injection \SCly\ coupled to DSA where some fraction of
shock-heated thermal particles re-cross the subshock and gain additional energy; \\
$\bullet$ shock-smoothing where the pressure from  superthermal particles in the shock precursor slows and heats the incoming plasma upstream of the viscous subshock; \\
$\bullet$ CR escape at an upstream \FEB\ (FEB);  \\
$\bullet$ a determination of the overall shock compression ratio $\Rtot$, taking into account escaping CRs and the modification of the equation of state from the production of \rel\ particles; \\
$\bullet$ fluctuating magnetic fields simultaneously calculated from resonant, short-, and long-wavelength instabilities generated from the CR current and pressure gradient in the shock precursor; \\
$\bullet$ momentum and position dependent particle diffusion calculated from  the self-generated magnetic fields; \\
$\bullet$ a consistent determination of the local scattering center speed  relative to the bulk plasma; and, \\
$\bullet$ \NL\ spectral energy transfer (i.e., cascading) and
dissipation of wave energy into the background plasma.
All of these processes are coupled together making a reasonably
consistent, albeit complicated, model.

In all that follows, the bulk plasma flow is in the positive
$x$-direction with speed $u(x)$ (see Figure~\ref{fig:Prof_M_En}).
The unmodified (i.e., far upstream) shock speed is $u_0$ and the downstream plasma speed,
in the shock frame, is $u_2=u_0/\Rtot$. In modified shocks,
there is a well-defined subshock compression $\Rsub=u_1/u_2$, where
$u_1$ is the plasma speed immediately upstream of the viscous
subshock. Everywhere, the subscript 0 (2)
implies far upstream (downstream) values.

\subsection{Small Angle Scattering} \label{sec:pitch}
The position and momentum dependent scattering mean  free path,
$\Lmfp(x,p)$, of a particle is determined from the local diffusion
coefficient $D(x,p)$.
A particle moves for a time $\deltime$ and then scatters
elastically and isotropically in the local frame through a small
angle $\thetaScat \le \sqrt{6 \deltime/t_c}$, where $t_c=\Lmfp/v_p
\gg \deltime$ \citep*[see][]{EJR90}. At this new $x$-position, the particle will have
changed momentum because of the converging bulk flow. For the next
$\deltime$, $\Lmfp$ is updated from the new local $D(x,p)$ and the
scattering process is repeated: the particle executes a random walk
on a six-dimensional sphere in space and momentum in an ever
changing magnetic and bulk flow background. This continues
 until the particle leaves the shock either by convecting far downstream or escaping at the upstream
 FEB.

\subsection{Thermal Leakage Injection}
A well-defined, essentially discontinuous
subshock with a subshock compression ratio $\Rsub < \Rtot$ is always
present in our smooth-shock solutions.\footnote{As mentioned above,
we do not attempt to model very short-scale Weibel-type
instabilities.}
As thermal particles injected upstream cross the subshock and
scatter in the downstream flow they gain enough energy so $v_p >
u_2$ for some  fraction of the population depending on the shock
Mach number. By virtual of random scatterings in the downstream
region, some $v_p > u_2$ particles will re-cross the
subshock into the precursor, gain additional energy, and enter the
first-order Fermi acceleration process.
This simple form of thermal leakage injection assumes the subshock is
transparent and ignores the possible existence of a
cross-shock potential or other effects from large amplitudes waves
that may occur at the subshock \citep[see, for example,][ for PIC
results showing the shock transition region]{kt10}.

The injection scheme
requires no superthermal seed particles and is entirely defined within the \mc\ scattering assumptions of Section~\ref{sec:pitch}.
Since the subshock strength and the downstream flow speed (via $\Rtot$) are determined \SCly\ with the global shock properties, the injection rate is coupled to DSA through the conservation conditions we discuss below.
No ``injection parameter" is needed to model injection from the thermal population.

\subsection{Mass-Energy-Momentum Conservation}
In our steady-state, plane-parallel shock, the conservation of mass flux is given by
\begin{equation}\label{massFlux}
\rho(x)u(x) = \rho_{0}u_{0}
\ ,
\end{equation}
where  $\rho(x)$
is the plasma density and $\rho_{0}u_{0}$ is the far upstream
mass flux.
The momentum flux conservation is given by
\begin{equation}\label{momentunFlux}
\PhiP(x)+P_{w}(x)=\PhiPz
\ ,
\end{equation}
where $\PhiP(x)$ is the particle momentum flux,
$P_{w}(x)$ is the momentum flux carried by the magnetic waves,
and $\PhiPz$ is the far upstream momentum flux, i.e., upstream from the free escape boundary where the interstellar magnetic field is
$\Bism$.\footnote{In our scenario, the ISM field consists of two components. There is a homogeneous part, $B_0$, and a turbulent part, $\Bkolm$, where the ISM turbulence is assumed to have a Kolmogorov spectrum. The total $\Bism^2 = B_0^2 + \Bkolm^2$.}

Separating the contributions from the thermal and accelerated
particles we have
\begin{equation}\label{momentunFlux1}
\rho(x)u^{2}(x) + \Pth(x) + \Pcr(x) + \Pw(x)=\PhiPz
\ ,
\end{equation}
where $\Pth(x)$ is the thermal particle pressure and $\Pcr(x)$ is
the accelerated particle pressure. A particle is ``accelerated" if it has crossed the subshock more then once and even though we use the subscript ``CR", the vast majority of accelerated particles will always be \nonrel.
Of course, if the acceleration is efficient, a large fraction of the pressure may be in \rel\ particles.

The energy flux conservation law is
\begin{equation}\label{momentunEnergy}
\PhiE(x)+F_{w}(x)=\Phi_{E0}
\ ,
\end{equation}
where $\PhiE(x)$ and  $F_{w}(x)$ are the energy fluxes
in particles and magnetic field correspondingly, and $\PhiEz$  is the energy flux  far upstream.
Taking into account particle escape at an upstream FEB, this can be re-written as
\begin{equation} \label{momentunEnergy1}
\frac{\rho(x)u^{3}(x)}{2} + \Fth(x) +
\Fcr(x) + \Fw(x) + \Qesc =
\PhiEz
\ ,
\end{equation}
where $\Fth(x)$ is the internal energy flux of the background
plasma, $\Fcr(x)$ is the energy flux of accelerated
particles, and  $\Qesc$ is the energy flux of particles that escape at the upstream FEB (note that $\Qesc$ is defined as positive even though CRs escape moving in the negative $x$-direction).

As seen in Figure~6 in \citet{EMP90} or Figure~8 in \citet{veb06} or Figure~2 in \citet{CS2013}, to give just three examples, the separation between ``thermal" particles and ``accelerated" particles in a shock undergoing DSA is not necessarily well defined.
Furthermore, energy exchange between the thermal and superthermal populations is certain to occur through non-trivial wave-particle interactions.
Nevertheless, the bulk of the plasma mass will always be in
quasi-thermal background particles and the internal energy flux of this background plasma can be
expressed as
\begin{equation}\label{F_th_P_th}
\Fth(x) =
u(x) \gamma_{g} \Pth(x)/(\gamma_{g} -1)
\ ,
\end{equation}
where $\gamma_{g}=5/3$ is the adiabatic index of the background
plasma.

\subsection{Magnetic Turbulence Generation and Dissipation}
\label{sec:TurbDiss}
We calculate the magnetic turbulence generated by the CR current, $\Jcr$, and pressure gradient, $d\Pcr/dx$, which simultaneously drive resonant, short--, and
long--wavelength instabilities.

\subsubsection{Energy Balance Equations} \label{sec:EnBal}
The spectral energy density of the magnetic
fluctuations $W(x,k)$, where $Wdk$ is the amount of energy in
the wavenumber interval $dk$ per unit spatial volume, can be calculated
including cascading. The
energy balance equation is
\begin{eqnarray} \label{eq:EnBal}
\frac{\partial \Fwc(x,k)}{\partial x} + \frac{\partial
\Pi(x,k)}{\partial k} &=&
u(x)\frac{\partial \Pwc(x,k)}{\partial x} + \nonumber \\
&& G(x,k) - \Lwc(x,k) \ ,
\end{eqnarray}
where $\Pi(x,k)$ is the flux of magnetic energy through $k$-space
towards larger $k$, and $G(x,k)$ and $\Lwc(x,k)$ are the spectral
energy growth and the dissipation rates, respectively.
The turbulence energy flux and pressure are given by
\begin{equation} \label{eq:Fw_x}
\Fw(x) = \int^{\kmax}_{\kmin}\Fwc(x,k) dk
\end{equation}
and
\begin{equation} \label{eq:Pw_x}
\Pw(x) = \int^{\kmax}_{\kmin}\Pwc(x,k) dk \ ,
\end{equation}
where the energy flux and pressure from the CR-current driven
fluctuations are given in equations~(\ref{F_W}) and (\ref{P_W})
below.
Equation~(\ref{eq:EnBal}) differs from equation~(1)
in \citet{vbe09} in that we now explicitly include adiabatic compression of the magnetic
turbulence energy density.

The energy flux and pressure from the CR-current driven
fluctuations that are included in equations~(\ref{momentunFlux1}) and (\ref{momentunEnergy1})
are defined as
\begin{equation} \label{F_W}
\Fwc(x,k) = u(x) \frac{\left( \left| \varphi(k) \right|^2 +2
\right)} {\left( \left| \varphi(k) \right|^2 +1 \right)} W(x,k) \ ,
\end{equation}
and
\begin{equation} \label{P_W}
\Pwc(x,k) = \frac{1}{\left( \left| \varphi(k)\right|^{2} +1\right)}
W(x,k) \ .
\end{equation}
For the case of CR-driven modes in a highly conducting plasma with
frozen-in magnetic fields, the velocity and field amplitudes  of a
harmonic perturbation are connected through the relation
\begin{equation} \label{eq:uk}
\delta\mathbf{u}_k =
\varphi(k) \frac{\delta\mathbf{B}_k}{\sqrt{4 \pi \rho}}
\ ,
\end{equation}
where $\varphi(k)$, with a complicated derivation, can be determined
from the dispersion equation~(\ref{SolveDispersMedium}) given
below and the mode polarizations.
For simplicity, we only present results for $\varphi(k) =1$
(as is the case for \alf\ modes), to be compared with those in
\citet[][]{vbe09}, but we have verified that the resulting spectra
of magnetic fields and particles do not vary significantly for the
range  $0.1 < |\varphi(k)| < 10$.

Integrating equation~(\ref{eq:EnBal}) from $\kmin$ to $\kmax$ we obtain
\begin{eqnarray} \label{eqvFw_k}
\frac{d\Fw(x)}{dx} &=&
u(x) \frac{d\Pw(x)}{dx} + \nonumber \\
&& \int^{\kmax}_{\kmin} G(x,k)dk - L(x)
\ .
\end{eqnarray}
The limits $\kmin$ and $\kmax$ are chosen to be outside the range containing a maximum scale determined by MFA and a minimum scale determined by dissipation. For these limits, the $\Pi(x,k)$ term in equation~(\ref{eq:EnBal}) vanishes.

For Kolmogorov-type cascading, the energy flux is  redistributed
over the energy scale but the total energy flux density, integrated
over wavenumbers is conserved.
Therefore, the total dissipation rate at any position $x$ is
\begin{equation}
L(x)=\int^{\kmax}_{\kmin}{\Lwc(x,k)}\, dk \ .
\end{equation}
We note that Kolmogorov-type  cascade models have been successful
at explaining the spectra of locally isotropic, incompressible
turbulence in non-conducting fluids observed in experiments and
simulations \citep[e.g.,][]{Biskamp2003}.
However, it is uncertain how spectral energy transfer operates in a
collisionless shock  precursor with a CR current strong enough to
modify the MHD modes, and in the presence of strong magnetic
turbulence.
In weak MHD turbulence, cascading was shown to be anisotropic
\citep[e.g.,][]{goldr97}: harmonics with wavenumbers  transverse to
the mean magnetic  field experience a Kolmogorov-like cascade, while
the cascade in wavenumbers parallel to the mean field is suppressed.

\subsection{Growth Rates of CR-Driven Instabilities}
\label{sec:Grow}
Self-generated magnetic turbulence is a key  ingredient for DSA and the production of
galactic cosmic rays
\citep[e.g.,][]{bell78,be87,JE91,md01,Parizot2006,
ber12,SchureEtal2012}.
Most analytical models that calculate MFA
\citep[e.g.,][]{bell04,plm06,Amato09,boe11,bbmo13} assume the
following ``homogeneous" form for the CR distribution function in the local rest frame of the upstream flow:
\begin{equation}\label{distrFuncCr}
f_{0}(\mathbf{p})= \frac{\nCR}{4\pi}N(p)\left[1+\frac{3
u_0}{c}\mu\right],
\end{equation}
where $\nCR$ is the concentration of CRs, $u_0$ is the shock
velocity,
$\mathbf{p}$ is the particle momentum in the plasma frame, 
$\mu = \cos{\theta}$, and $\theta$ is the particle pitch angle, i.e., the angle between $\mathbf{p}$ and the magnetic field  assumed to lie in
the $x$-direction. The CR distribution function, $N(p)$, is
normalized as
$\int_{\pmin}^{\pmax} N(p) p^2 dp = 1$,
where
$\pmin$ and $\pmax$ are the minimum and maximum particle
momenta in the CR distribution at the upstream position.
In analytic or semi-analytic treatments, $\nCR$ and $u_0$ are
normally assumed to be position independent and $N(p)$ is often
assumed to be a power law.

In the \mc\ code, the particle distribution function contains  more
information than represented by equation~(\ref{distrFuncCr}). It is
calculated directly in the modified shock precursor with a varying
$\nCR(x)$ and $u(x)$, rather than constant values, and  no
approximations are made restricting the particle anisotropy.
Instead of $f_{0}(\mathbf{p})$, we can use
\begin{equation}\label{distrFuncCr1}
\fcr(x,\mathbf{p})= \frac{\nCR(x)N(x,p)}{4\pi}+\frac{3
\JcrI(x,p)}{4\pi v_p}\mu \ ,
\end{equation}
to calculate the local growth rates. Here, $\JcrI(x,p)$ is the
differential CR current, $v_p$ is the particle velocity in
the local frame, and $N(x,p)$  is determined \SCly\ with the shock
structure and will not be a power law in the shock precursor.
The full CR current  (for now we consider only protons) is
\begin{equation} \label{CurrentCr}
\Jcr(x) = e \int_{\pmin(x)}^{\pmax}\JcrI(x,p)\, p^{2}dp
\ ,
\end{equation}
and this is used to calculate  the CR-driven instabilities.
Note that $\JcrI(x,p)$ only contains superthermal particles  and
$\pmin$ depends on $x$, increasing as the  observation position
moves further upstream from the subshock.

At any $x$-position in the precursor, we use local \mc\  values for
$\nCR(x)$, $u(x)$, and $\Jcr(x)$ to calculate the dispersion
relation, ignoring effects from spatial gradients in our ``local
homogeneous" approximation.
Specifically, the growth rates for the
three CR-driven instabilities are derived from the dispersion relation for modes
propagated along the mean magnetic field in the magnetized
background plasma.
This dispersion relation, which includes the short--wavelength Bell instability \citep[i.e.,][]{bell04},
was derived in \citet{boe11,bbmo13}, and is
\begin{eqnarray} \label{dispersMedium}
&& \omega^{2}\mp\omega
ik k_c \frac{\alpha_{t}}{4\pi\rho}\left[\frac{1}{2}\frac{eA(x,k)}
{\Jcr(x)}+
\frac{3}{2}\right]-k^{2}v_{a}^{2}\pm
\nonumber \\
&&
\pm k k_c v_{a}^{2}\left(1+\frac{
\kappa_t}{\Bls(x,k^{*})}\right)\left[\frac{eA(x,k)}{\Jcr(x)}-1\right]
= 0,
\end{eqnarray}
where $\rho$ is the background plasma density, and the $\pm$ signs,
here and in the equations that follow, correspond to the two
circularly polarized modes. The mean magnetic field, $\Bls(x,k)$, in the
dispersion relations is defined as the sum of the long-wavelength
(i.e., large-scale ``ls") harmonics plus the ambient homogeneous field, $B_0$, i.e.,
\begin{equation}\label{B_ls}
\Bls(x,k) = \sqrt{4\pi\int_{\kmin}^{k}W(x,k^{'})dk^{'}+B_{0}^{2}} \
.
\end{equation}
The definitions for $A(x,k)$ (equation~\ref{coeffANu})
and $k^{*}$ (equation~\ref{eq:kstar}) are given below and $\alpha_t$ and $\kappa_t$
are defined in Section 5.1 of \citet{boe11}.

The solution to equation~(\ref{dispersMedium}) is
\begin{equation} \label{SolveDispersMedium}
\omega = \left(\pm\sqrt{d^{2}+4 b} -d\right)/2
\ .
\end{equation}
In equations~(\ref{dispersMedium}) and (\ref{SolveDispersMedium}),
\begin{equation} \label{dCoeff}
d = \mp
ik k_c \frac{\alpha_{t}}{4\pi\rho}\left[\frac{1}{2}
\frac{eA(x,k)}{\Jcr(x)}+\frac{3}{2}\right],
\end{equation}
\begin{equation} \label{cCoeff}
b = k^{2}v_{a}^{2}\left[1 \mp \frac{k_c}{k}\left(1+\frac{
\kappa_{t}}{\Bls(x,k^{*})}\right)\left(\frac{eA(x,k)}{\Jcr(x)}-1
\right)\right],
\end{equation}
\begin{equation}\label{coeffANu}
A(x,k)=\int_{0}^{\infty}\sigma(p)\JcrI(x,p)p^{2}dp
\ ,
\end{equation}
\begin{eqnarray}\label{sigmaCrNu}
& &
\sigma(z) =
\frac{3}{2z^{2}}+\frac{3}{8z}\left(1-\frac{1}{z^{2}}+\left(\frac{a}{z}\right)^{2}\right)\Psi_{1}
-\frac{3a}{2z^{3}}\Psi_{2}
\nonumber \\
 &\mp& i\left\{\frac{3}{4z}\left(1-\frac{1}{z^{2}}+\left(\frac{a}{z}\right)^{2}\right)\Psi_{2}
 -\frac{3a}{2z^{2}}+\frac{3a}{4z^{3}}\Psi_{1}\right\},
\\
  & &
\Psi_{1}(z) =
\ln\left[\frac{(z+1)^{2}+a^{2}}{(z-1)^{2}+a^{2}}\right]\, ,
 \nonumber \\
\mathrm{and,}
 \nonumber \\ \nonumber \\
 & &
\Psi_{2}(z) =
\arctan\left(\frac{z+1}{a}\right) +
\arctan\left(\frac{z-1}{a}\right)
\ .
\nonumber \end{eqnarray}
In these equations, $v_a = \Bls(x,k)/\sqrt{4 \pi \rho(x)}$, $z =
kcp/(e \Bls)$, $a$ is the collision parameter equal to the ratio of
the gyroradius of a CR particle to its mean free path determined by
scattering on magnetic fluctuations, and $k_c = 4 \pi |\Jcr(x)|/[c
\Bls(x,k)]$ is Bell's critical wavenumber at position $x$ in the
precursor.

The effective, self-generated,  magnetic field is given by
\begin{equation} \label{eq:Beff}
\Beff(x) =\sqrt{4\pi\int_{\kmin}^{\kmax}W(x,k^{'})dk^{'}} ,
\end{equation}
or equivalently,

\begin{equation} \label{eq:Beff2}
\Beff(x) = \sqrt{\Bls^{2}(x, \kmax)- B_{0}^{2} }
\ .
\end{equation}
The total field is $\Btot=\Bls(x,\kmax)$ so $\Btot^2 = \Beff^2 + B_0^2$. As mentioned above, we assume the ambient ISM field consists of a uniform component $B_0=3$\,\muG, and a turbulent part generated from the background \Kol\ turbulence assumed to have a strength such that $\Bkolm=3$\,\muG. Therefore, for the ISM, $\Btot \simeq 4.2$\,\muG.

\subsubsection{Long--wavelength Instability}
The plasma fluctuations created by Bell's short--wavelength
instability influence the plasma dynamics and this  effect is
modeled with the ponderomotive coefficients, $\alpha_{t}$ and
$\kappa_{t}$, in equations~(\ref{dCoeff}) and
(\ref{cCoeff}).\footnote{If $\alpha_t=\kappa_t=0$,
equation~(\ref{dispersMedium}) gives the standard current driven
resonant and Bell instabilities.}
These ponderomotive effects result in the long--wavelength plasma
instability (LWI) developed  by \citet*{boe11} and \cite{bbmo13}.

The ponderomotive coefficients are
determined by the mean square of the short-scale magnetic
field fluctuations produced by Bell's instability, i.e.,
\begin{equation}
\kappa_{t}/[\Bls(x,k^{*})] = \pi N_B
\ ,
\end{equation}
and
\begin{equation}
\frac{k_0 \alpha_t}{4 \pi \rho} =
2 \pi \sqrt{\xi} N_B v_a
\ ,
\end{equation}
where $\xi$ is the dimensionless mixing length of the short-scale
turbulence as defined in \citet{boe11}, $v_a$ is the \alf\ speed which is the characteristic speed of the medium,  and $N_B$, the dimensionless amplitude of the magnetic fluctuations, is defined below.
While $N_B$  is determined in the \mc\ simulation from the CR
distribution, $\xi$ is not. In the simulations reported here, we
take $\xi = 5$ but note  that our results are only weakly dependent
on $\xi$.

To achieve  a solution, we set the ponderomotive coefficients  in
the dispersion equation~(\ref{SolveDispersMedium}) equal to  zero
for wavenumbers $k>k^{*}$, where $k^{*}$ is determined by the
effective resonant condition
\begin{equation} \label{eq:kstar}
k^{*}c \, \pmin (x) = e \Bls(x,k^{*})
\ .
\end{equation}
At any $x$-position we set  $N_B(x)$ = 0 for $k^{*}(x)> k_c (x)$,
where there is no wave growth from Bell's mode,
and we set
\begin{equation} \label{N_B}
N_B(x) =
\frac{\sqrt{\Bls^{2}(x,k_c) - \Bls^{2}(x,k^{*})}}{\Bls(x,k^{*})}
\ ,
\end{equation}
for $k^{*}(x) < k_c(x,k^{*})$, where Bell's instability operates.
The mode growth rates, $\Gamma(x,k)$, in the turbulence energy  balance equation~(\ref{eqvFw_k}), where $G$ is the energy
growth rate (see equation~\ref{eq:SpEnDiscrit} below),  are connected to the
roots of the dispersion equation as
\begin{equation}
\Gamma(x,k) = 2\, \mathrm{Im}[\omega(x,k)]
\ ,
\end{equation}
where the 2 accounts for the fact that the energy in
turbulence  is proportional to the square of the amplitude of
$\Delta B$.
At each $x$-position  we choose the mode with the maximum value of
$\mathrm{Im}[\omega(x,k)]$, if positive.

Because we use the full, anisotropic CR distribution function in
equation~(\ref{dispersMedium}) and the substitutions that follow,
the dispersion equation~(\ref{SolveDispersMedium})
simultaneously accounts for the CR-pressure driven resonant
streaming instability, and the two CR-current driven
instabilities.
To our knowledge, this is the first attempt to combine
these instabilities in a consistent, broadband
shock model.\footnote{We note
that we do not include the ion-acoustic instability
\citep[e.g.,][]{df86,md01} which may be important for plasma heating
in the precursor.}

\subsubsection{Dissipation and Fluxes of Particles
and Waves}\label{sec:FxPWave}
The self-generated turbulence is assumed to suffer viscous dissipation at a rate
proportional to $k^2$, and the dissipated energy is pumped directly
into the thermal particle background.
The background plasma energy balance is governed  by the plasma
compression and the turbulence dissipation rate
and obeys the equation
\begin{equation} \label{eqvFth}
\frac{d\Fth(x)}{dx} = u(x) \frac{d \Pth(x)}{dx} + L(x) \ .
\end{equation}
If the heating of the background plasma by the
turbulence dissipation is weak
[i.e., $L(x) \sim 0$],
equations~(\ref{massFlux}), (\ref{F_th_P_th}), and (\ref{eqvFth})
result in the adiabatic compression of the background plasma.

The results of \citet{vbe08},
which included only the
resonant CR-streaming instability for MFA with a variable dissipation rate,
demonstrated that even a modest
rate of turbulence dissipation can significantly increase the
precursor temperature (see Figure~\ref{fig:Temp} below) and that this, in turn, can increase the rate
of injection of thermal particles \citep[see also][]{vbe09}.
However, the \NL\ feedback of these
changes on the shock structure tend to cancel so that the
spectrum of high energy particles is only modestly affected.
As described in \citet{vbe09}, we only apply dissipation to our models with cascading.

The relation between the CR pressure gradient and the CR energy
flux $\Fcr$, can be written as
\begin{equation} \label{eqvFcr}
\frac{d\Fcr(x)}{dx} =
\left[u(x) + \Vscat(x)\right] \frac{d \Pcr(x)}{dx}
\ ,
\end{equation}
where $\Vscat(x)$ is the scattering center velocity measured in the local frame.\footnote{Note that in the precursor $\Vscat$ will always be negative, that is directed upstream in the local frame.}
This definition is a position dependent generalization of the
position independent  ``wave frame" velocity  introduced by
\citet{skilling71} for CR interactions with hydromagnetic waves.
An important
feature of the CR-current driven modes is that, in the most
interesting cases with  $\mathrm{Im}[\omega(k,x)] > k v_{a}$, the
scattering center velocity must be substantially less than the \alf\
speed $\Valf$ in order to achieve energy conservation. This is true whether $B_0$ or $\Bls$ is used to calculate
$\Valf$ and is discussed in detail in \S\ref{sec:ScatV}.

As follows from equation~(\ref{F_W}), the energy flux $\Fwc(x,k)$ may
span the range from $W(x,k)$ to $2\,W(x,k)$, for $\varphi(k)$
between 0 and 1. In our \mc\ simulations, we use $\varphi(k)=1$ but
our results are not sensitive to $\varphi(k)$. For resonantly
generated \alf\ modes, \cite{McKVlk82} derived
\begin{equation}
\Fwc(x,k) = \left[3u(x)/2 + \Vscat(x) \right] W(x,k)
\end{equation}
and
\begin{equation}
\Pwc(x,k) = W(x,k)/2 \ ,
\end{equation}
which corresponds to $\varphi(k) = 1$, since  the kinetic and
magnetic energy densities are exactly equal for  \alf\ modes.

\subsection{Effective Scattering Center Velocity} \label{sec:ScatV}
The turbulence produced by shock accelerated particles can move
relative to the bulk plasma and this movement must be \SCly\
included when determining the \NL\ shock structure. If the
turbulence is assumed to be \alf\ waves, the scattering center speed
can be taken to be the \alf\ speed with the far upstream field,
$v_{a0}(x) = B_0 / \sqrt{4 \pi \rho(x)}$,  and the turbulence can be
calculated in equation~(\ref{eqvFw_k}) in a straightforward fashion
using
\begin{equation} \label{v_a_dP_cr}
\int^{\kmax}_{\kmin} G(x,k)dk = -v_{a0}(x) \frac{d\Pcr}{dx} \ ,
\end{equation}
to model the wave growth.

Equation~(\ref{v_a_dP_cr}) was obtained by \citet{McKVlk82} assuming
(i) that the resonant wave-CR particle interactions are
quasi-linear with $\lambda \propto 1 / W(x,k)$, and
(ii) that the magnetic turbulence is dominated by \Qlin\ modes with
growth rates $\Gamma(x,k) \ll k v_{a0}$ (for simplicity we only
consider modes propagating parallel to the mean magnetic
field), and (iii) that the mode growth rate is a linear function of the isotropic part of 
the distribution function $f(x,p)$.
One ambiguity with this approach is the choice of $B$.  It is
typically chosen as either $B_0$ or some fraction of the amplified
magnetic field at $x$.

In considering equation~(\ref{v_a_dP_cr}), however, it must be noted
that non-resonant, CR current-driven modes
\citep[e.g.,][]{bell04,boe11,SchureEtal2012,bbmo13} have their
fastest growth rates when $\Gamma(x,k) > k v_{a0}$ and these modes
dominate the magnetic fluctuation spectra.
This point can be illustrated in a simple way. Consider just the  growth rate of Bell's instability without the LWI.
Then, the solution to equation~(\ref{dispersMedium}) is
\begin{equation}\label{poyasnenye}
\omega = \pm\sqrt{v_{a0}^{2}k^{2} + K(k,\JcrI)} \ ,
\end{equation}
where $K(k,\JcrI)$ is determined by the CR current. This equation can
be obtained directly from equation~(\ref{SolveDispersMedium}) by setting
$\alpha_{t} = \kappa_{t}=0$, the two parameters  responsible for the
LWI.
To get equation~(\ref{v_a_dP_cr}), two conditions must hold.  The first
is that  $\left|K\right|\ll v_{a}^{2}k^{2}$ and therefore the square
root in equation~(\ref{poyasnenye}) can be expanded as
\begin{equation}  \label{poyasnenye1}
\omega\approx\pm v_{a0}k\left(1+\frac{K}{2v_{a0}^{2}k^{2}}\right).
\end{equation}
The second condition is that $\Gamma(x,k)$ is a linear function of
the distribution function $f(x,p)$. Note that $G(x,k) = \Gamma(x,k)
W(x,k)$ is generally assumed in our approach.

While both of these
conditions can be fulfilled in the case of weakly growing
\alf-like turbulence, weak growth is not expected
if the CR current is as large as
believed to occur in
young SNRs.
If the CR current is large, and non-resonant current driven
instabilities are important, then $\left|K\right|\gg
v_{a0}^{2}k^{2}$  and equation~(\ref{poyasnenye}) simplifies to
\begin{equation}\label{poyasnenye2}
\omega\approx\pm\sqrt{K}
\ ,
\end{equation}
that is, $\omega$ is proportional to the square of $K$ rather than
proportional to $K$. While $K(k,\JcrI)$ is a linear function of $f(x,p)$ 
(if the diffusion approximation is valid), it is clear from
equation~(\ref{poyasnenye2}) that the wave growth term $G(x,k)$ is a
\NL\ function of $f(x,p)$ for large CR currents.

We believe this reevaluation of the  self-generation of turbulence
by non-resonant modes is fundamentally important. The non-resonant
turbulence has a character quite different from \alf\
waves\footnote{We note that the turbulence found in PIC and hybrid simulations
\citep[e.g.,][]{kt10,CS2014}, as well as turbulence generated near
interplanetary shocks \citep[e.g.,][]{BOEF97,KajdiEtal2012}, is
not necessarily well described as \alf\ waves.}
and we find that equation~(\ref{v_a_dP_cr}), regardless  of the
choice of $B$, does not allow for momentum and energy conserving
solutions.
With the \mc\ technique, we obtain consistent solutions by simply generalizing
equation~(\ref{v_a_dP_cr}) to
\begin{equation} \label{VscatGradPcr}
\Vscat(x)\frac{d \Pcr}{d
x}=-\int^{\kmax}_{\kmin} G(x,k) dk
\ ,
\end{equation}
assuming there is a single, position dependent, effective scattering
speed, and including $\Vscat(x)$ in our iterative scheme.
Equation~(\ref{VscatGradPcr}) is derived using
equations~(\ref{massFlux}), (\ref{momentunFlux1}),
(\ref{momentunEnergy1}), (\ref{eqvFw_k}), (\ref{eqvFcr}),  and the
 equation of state of the background plasma.

At each iteration, the $i$-th$+1$ value of the scattering center
velocity, $\VscatI(x)$, is obtained from
\begin{equation} \label{Vscat}
\VscatI(x) = -\frac{\int^{\kmax}_{\kmin}{\Gamma^{(i)}(x,k)
W^{(i)}(x,k)}dk}
{d\Pcr^{(i)}/dx}
\ ,
\end{equation}
where $\Gamma$, $W$, and $d\Pcr/dx$ are all determined in the  \mc\
simulation in the $i$-th iteration. With equation~(\ref{Vscat})
there is no need to associate $\Vscat$ with the \alf\ speed.
In the upstream region, accelerated particles are propagated
through the shock assuming the scattering center velocity is $u(x) +
\Vscat(x)$, while the velocity is $u(x)$ for
the background thermal particles. Downstream from the shock, we take
$\Vscat=0$.

An essential and unique element of our calculation is that when  the
integral in equation~(\ref{VscatGradPcr}), with $\Vscat$ determined
from equation~(\ref{Vscat}), is used to replace the integral in the
energy balance equation~(\ref{eq:EnBal}), we make use of the full
anisotropic CR distribution function including the pressure gradient
and the CR current.
Our derivation of $\Vscat(x)$ accounts for the anisotropic magnetic
modes  with dispersive properties (phase and group velocities)
determined by both the background plasma and the CR angular and
momentum distributions.
As we show below  (see Figure~\ref{fig:vscat}), while the fastest growing CR-driven modes are highly anisotropic,
their phase and group velocities are typically
strongly \subAlf\
with $\Vscat(x)\ll u(x)$ for all $x$.

Significantly, even though $\Vscat(x)$ may be small,  it has a
strong influence on particle acceleration and must be taken into
account to determine a consistent shock structure in the \mc\ model.
Despite its wide use for many years, we caution that replacing the
wave-growth term in equation~(\ref{eqvFw_k}) with
equation~(\ref{v_a_dP_cr})
is a poor approximation when short-- and long--wavelength
instabilities are taken into account.

\subsection{Particle Mean Free Paths}
The \mc\ code determines $\Lmfp(x,p)$ using
various analytic approximations and the locally averaged magnetic
field. As indicated in Figures~\ref{fig:fp_mfp_DS}
and \ref{fig:fp_mfp_UpS}, we define different regimes for determining $\Lmfp$ starting with
thermal particles \citep[see][for additional details]{vbe09}.

\subsubsection{Thermal Particles}
Thermal particles enter the simulation upstream of the subshock at a
position which depends on the plasma flow velocity gradient and then
propagate toward the subshock.
Far
upstream, and during the first iteration when the shock is
unmodified and before MFA has generated turbulence in the precursor,
these thermal particles with momenta $\pth$ experience relatively weak turbulence and have a
relatively large $\Lmfp=\rg$, where $\rg \sim \pth c/(e \Bls)$.

It subsequent iterations, when strong turbulence  exists in the
precursor, the $\Lmfp$ for low energy particles can become
considerably smaller. At this point, we take into account the fact
that the transport of low-energy particles may not be diffusive but
can be governed by turbulent advection of particles frozen into
large-scale turbulent plasma vortexes \citep[see][ for more
details]{bt93}.

\subsubsection{Vortex Advection for Low-Energy Particles}
\label{sec:vortex}
The \mc\ simulation describes turbulence  on
mesoscopic scales. At the low end of this
mesoscopic range, turbulence, particularly if produced with a
quasi-power-law spectrum by cascading, may result in low-energy
particles having mean free paths smaller than the scale of vortexes
that are expected to develop. In this case, the low-energy particles
can be `trapped' by the vortexes and execute non-diffusive
transport. We have developed a transport model that mimics the
essential physics for vortex advection  in and near the viscous
subshock layer and this is included in our model.

Following the discussion in \citet{Vladimirov09dis}, we  assume that
low-energy particles can be confined by resonant scattering and
trapped within turbulent plasma vortexes of different scales.
In this case, the transport of these particles on  scales greater
than the correlation length of the turbulence (i.e., greater than
the largest turbulent harmonics), is governed by the turbulent
advection of the vortexes rather than by resonant diffusion of
individual particles.
This vortex trapping mimics how low-energy particles would interact with local displacements and distortions of a shock front moving through the turbulent ISM.
If the Bohm diffusion coefficient is small enough, particles can be considered to be `anchored' to a small section of the shock front and it was shown by \citet{b82} that there is a `diffusion regime' in space and time of the average displacement of a section of the subshock surface. The low-energy particles anchored to this section have an effective transport that is diffusive.

Our recipe for the turbulent transport of low-energy particles can be summarized as follows:
%The results of \citet{bt93} can be summarized as follows:  
the particle diffusion coefficient due to turbulent vortex
advection, $\Dvor(x)$, is momentum independent and determined by
\begin{equation} \label{D_wh}
\Dvor(x) = \Uvor(x)\, \Lvor(x)
\ ,
\end{equation}
where $\Uvor$, the typical speed of turbulent motions with
correlation length $\Lvor$, is estimated from
\begin{equation} \label{U_wh}
\Uvor(x) =
\sqrt{\frac{\int_{\Kvor(x)}^{\kmax}W(x,k^{'})dk^{'}}{\rho}}
\ .
\end{equation}
Here, $\Kvor(x)=2\pi \Lvor^{-1}(x)$ and $\kmax$ is determined by
the turbulence dissipation mechanism.
For concreteness we take
\begin{equation}
\Lvor(x) = 0.5\, |x|
\ ,
\end{equation}
and determine the \mfp\ from vortex motion from
\begin{equation} \label{eq:Lvor}
\LamVor(x) = 3 \Dvor(x)/v_p
\ .
\end{equation}
This `convective' diffusion coefficient is indicated with the label `1' in Figures~\ref{fig:fp_mfp_DS} and \ref{fig:fp_mfp_UpS}.

For `trapped' low-energy
particles, $\Dvor(x)$ can be much greater
than the resonant scattering coefficient.
Low energy CRs in the vicinity of the subshock have a high frequency of scattering and because of this they are tied to the
large-scale vortexes. Their transport is dominated by the convection of the turbulence instead of microscopic scattering off waves.
At every \sas\ event, we  compare the \mfp\ from
magnetic fluctuations, $\lambda(x,p)$ (discussed in
Section~\ref{sec:mfp} below),  to $\LamVor(x)$.
The larger of the two is used to propagate the low-energy particles.
This will modify the injection
process but injection will still be \SCly\ determined in the \mc\
simulation as the shock structure adjusts to conserve momentum and energy. Our results are not strongly dependent on the scattering assumptions made for low-energy particles.

\subsubsection{Particle Scattering by Short-Scale Fluctuations}
The wave number $\Kres$ associated with resonant interactions
of particles with  momentum $p$ is given by
\begin{equation}
\frac{\Kres cp}{e \Bls(x,\Kres)} = 1 \ .
\end{equation}
Our model includes the short-scale turbulence produced by
CR-driven instabilities where the wave number of the short-scale modes $\Kss \gg \Kres$ and the modes are defined by
\begin{equation}
\frac{\Kss cp}{e \Bls(x,\kmax)} \gg 1 \ .
\end{equation}
For particles with gyroradii $\rg=cp/(e \Bls)$ much  larger than the
scale-length of the short-scale turbulence, the mean free path produced by the short-scale modes is
\begin{equation} \label{MFP_s}
\Lss(x,p) =
\frac{4}{\pi}\frac{\Rss^{2}}{\Lcor}
\propto p^2
\ ,
\end{equation}
where $\Rss = cp/(e \Bss)$,
\begin{equation}\label{B_ss}
\Bss(x,\Kres) = \sqrt{4\pi\int_{\Kres}^{\kmax}W(x,k^{'})\, dk^{'}} \
,
\end{equation}
and
\begin{equation} \label{l_cor}
\Lcor = \frac{\int_{\Kres}^{\kmax} [W(x,k^{'}) /k^{'}] \, dk^{'}}
{\int_{\Kres}^{\kmax}W(x,k^{'})dk^{'}} \ .
\end{equation}
It has been shown that the short-scale scattering regime with $\Lss
\propto p^2$ holds even for large amplitude magnetic field
fluctuations \citep[see, for example,][]{jokipii71,topt85}. This
mode is particularly important because it dominates  particle
scattering for the highest energy CRs a given shock can produce.

\subsubsection{Effective Particle Mean Free Path} \label{sec:mfp}
The total effective scattering \mfp\ is
\begin{equation}
\lambda(x,p)=\max\{\LamVor(x),\lambda_{s}(x,p)\}
\ ,
\end{equation}
where
\begin{equation} \label{eq:EffMFP}
\lambda_{s}(x,p)=\frac{1}{\Lpic^{-1}(x,p)+\Lss^{-1}(x,p) +
\Lres^{-1}(x,p) + \Lcor^{-1}}.
\end{equation}
The diffusion regimes included in equation~(\ref{eq:EffMFP}) are consistent with those seen in numerical simulations of particle transport in strong magnetic turbulence 
\citep[e.g.,][]{casse02,marcowith06,reville08}.

For low-energy particles we define a  transition momentum $\pTran =
\fTran m_{p} u_0$ where  $\fTran > 1$  is a free parameter  and $\Lpic(x,p)$ is the Bohm
diffusion length, defined here by\footnote{The subscript `pic' suggests that $\Lpic$ can be determined from PIC simulations.}
\begin{equation}
\Lpic(x,p)=\frac{cp}{e \Bls(x,\Kres)} .
\end{equation}
For particles with  momenta $p> \pTran$, the mean free path is
determined by
\begin{equation}
\Lpic(x,p) = \frac{c \pTran}{e \Bls(x,\Kres)}
\left(\frac{p}{\pTran}\right)^{2}.
\end{equation}%
In the simulations reported here, we
take $\fTran$ = 3.0.

The mean free path due to quasi-resonant scattering is
\begin{equation} \label{eq:Lres}
\Lres(x,p)=\frac{1}{\pi^2}\frac{cp \Bls(x,\Kres)}{e \Kres
W(x,\Kres)} ,
\end{equation}
and  $\Lcor$ is defined as
$\min[\Lls(x,p),\Kls^{-1}(x,p)]$ where
\begin{equation}
\Lls(x,p)=\frac{\int_{\kmin}^{\Kres} [W(x,k^{'}) /k^{'}] \, dk^{'}}
{\int_{\kmin}^{\Kres}W(x,k^{'})dk^{'}}.
\end{equation}
Here $\Kls$ is the maximum wavenumber satisfying the relation
$\displaystyle \Kls W(x,\Kls)>\int_{\kmin}^{\Kls}W(x,k^{'})dk^{'}$,
for $\Kls <\Kres$. If $\displaystyle \Kls
W(x,\Kls)<\int_{\kmin}^{\Kls}W(x,k^{'})dk^{'}$ for any $\Kls$ in the
range $\kmin < \Kls <\Kres$, then $\Kls =\kmin$.

The parameter $\fTran$ is required because the \mc\ technique
does not model production of the short-scale turbulence responsible
for formation of the viscous subshock, i.e., that produced by the Weibel instability.
While PIC simulations can do this,  they cannot yet cover the wide
dynamic range needed to produce the high-energy CRs responsible for
the large-scale turbulence and cascading that produce the vortex
transport we described in Section~\ref{sec:vortex}. Parameterizing
the transition between subshock and vortex advection is currently
the best way to describe the transport of superthermal particles
until they reach $p > \pTran$ and equation~(\ref{eq:EffMFP}) can be used.
We note that our results are sensitive to $\pTran$ for low shock
speeds ($u_0 \sim 1000$\,\kmps) but become less sensitive as $u_0$
increases.

\newlistroman

\subsection{Monte Carlo Iterative Solution} \label{sec:MC}
The coupled components of our steady-state, plane-shock simulation are
solved iteratively. Given that the particle acceleration process generates full particle spectra $f(x,p)$, CR currents,
and CR pressure gradients, at all positions relative to the
subshock, we can write the spectral energy density at the $i$-th iteration, $W^{(i)}(x,k)$, as\footnote{While the energy flux
of escaping CRs is included \SCly, we do not calculate here the
turbulence generated by this flux. In efficient DSA, the escaping
flux at the upstream FEB is produced by the highest energy CRs and
the turbulence generated by these CRs may significantly influence
the maximum CR energy the shock can produce. Work to account for the
turbulence generated by escaping CRs is in progress.}
\begin{eqnarray} \label{eq:Witer}
&& u^{(i-1)}(x)\frac{\partial W^{(i)}(x,k)}{\partial x}+
\frac{3}{2}\frac{du^{(i-1)}(x)}{dx}W^{(i)}(x,k) + \nonumber \\
&+& \frac{\partial \Pi^{(i-1)}(x,k)}{\partial k} =
\Gamma(x,k,W^{(i-1)})W^{(i-1)}(x,k) + \nonumber \\ &-&
\Lwc(x,k)^{(i-1)} \ .
\end{eqnarray}
Equation~(\ref{eq:Witer}) is a discretized version of
equation~(\ref{eq:EnBal}) where we
assume that the small magnetic turbulence increment between any two
iterations can be estimated using the \Qlin\ growth rate of
the turbulence.
Then, the magnetic turbulence growth rate
\begin{equation} \label{eq:SpEnDiscrit}
G^{(i)}(x,k) = \Gamma[x,k,W^{(i-1)}] W^{(i-1)}(x,k)
\end{equation}
is derived using
the CR-current and the mean magnetic field
$\Bls^{(i-1)}(x,k)$ (which is
defined by equation~\ref{B_ls}) with $W^{(i-1)}(x,k)$ derived on
the $i$-th-1 iteration.
This approach is somewhat similar to mean field models used in the
statistical theory of ferromagnetism \citep[e.g.,][]{Kittel76},
and while it neglects some correlation
effects and we assume randomization of the field direction,
we contend it overcomes the major limitations
of \Qlin\ theory.
We use the \Qlin\ approximation but we only apply
equation~(\ref{eq:SpEnDiscrit}) between  iterations where $\Delta
B^{(i-1)} < \Bls^{(i-1)}$. This allows us to slowly increase $\Bls$
to values $\Bls \gg B_0$ which are currently beyond any exact
simulation method without violating $\Delta B < \Bls$ in any one
iteration step.

For a given set of shock parameters, we start with an unmodified
shock with  compression ratio, $\rRH$, determined by the \RH\
conditions, $\Vscat(x)=0$,
and $\mfpTH  = \rgTH = \pth c/(e B_0)$, where $\pth$ is the thermal particle momentum.
The initial magnetic turbulence is taken to be
\begin{equation} \label{Bibint}
W(x,k) =
\frac{B_{0}^{2}}{4\pi}\frac{k^{-5/3}}
{\int_{\kmin}^{\kmax}k^{-5/3}dk}
\ .
\end{equation}
Thermal particles are injected far upstream and diffusively
accelerated, leaving the shock by convecting far downstream or
escaping at the upstream FEB.
After the first iteration, $f(x, p)$ is determined, along with
$\Jcr(x)$ and $d\Pcr(x)/dx$, and these are used to calculate
$\Gamma(x, k)$ and $W(x, k)$ for the next iteration.
From these we determine
$F_{w}(x)$, $P_{w}(x)$, $D(x, p)$, and $\Vscat(x)$.
We include cascading and energy dissipation which
transfers energy from $W(x,k)$ to the background plasma influencing
the subshock strength and particle injection.

\begin{figure}
\epsscale{1.0}
\plotone{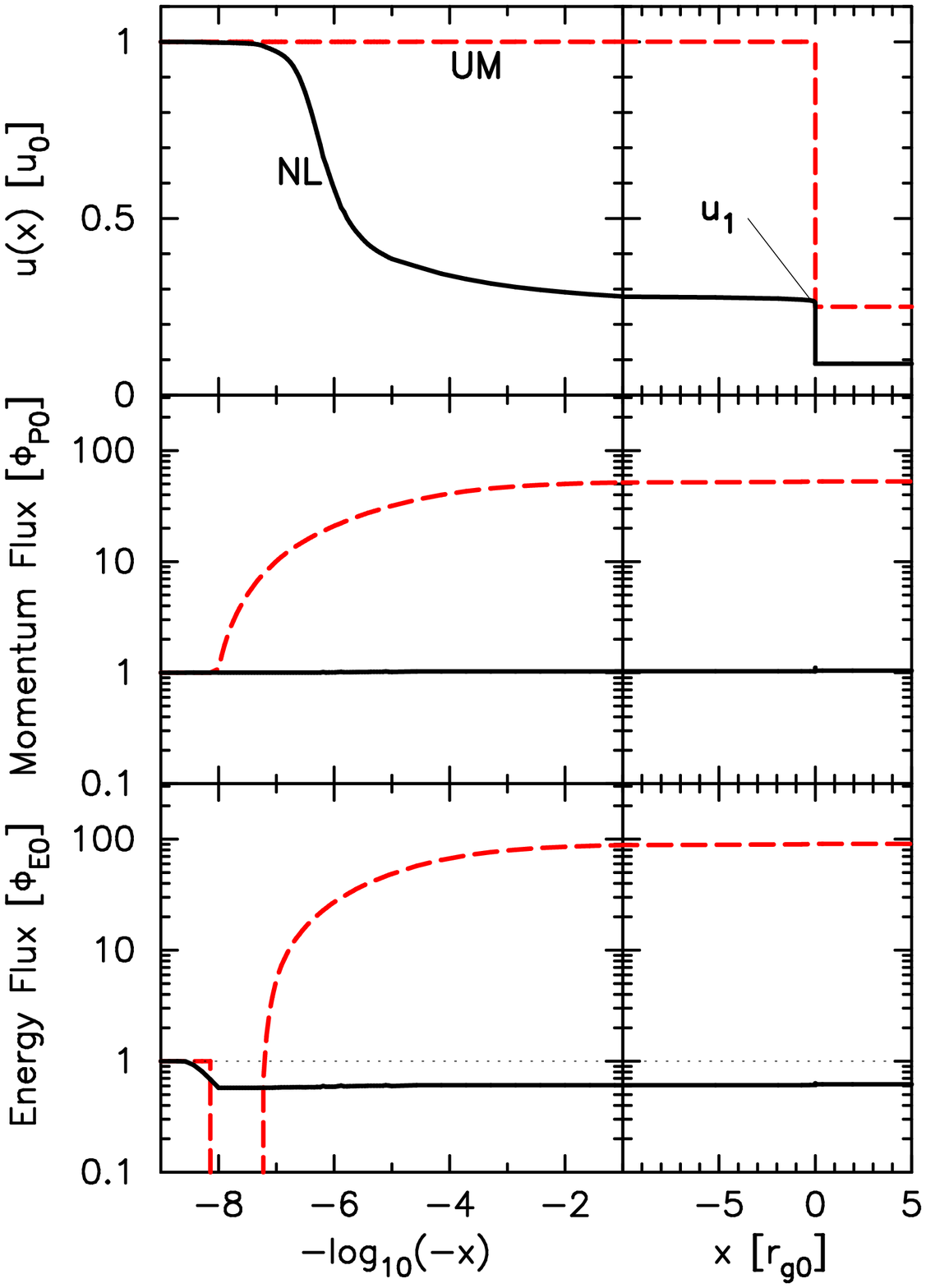}            % Fig 1
\caption{The dashed (red) curves show the results for an unmodified shock with $\Rtot \simeq 4$ (Model \aafourUM).
The solid (black) curves (Model \aafour) show the \SC\ result where the momentum and energy fluxes are conserved across the shock. For this example, where all three instabilities are active, the \SC\ compression ratio is $\Rtot \simeq 11.3$ and $\sim 40$\% of the energy flux is lost at the FEB at $x=-10^8\,\rgZ$, where $\rgz \equiv m_p u_0 c/(e B_0) \simeq 5.6\xx{-9}$\,pc. Both models have $u_0=5000$\,\kmps, $n_0=0.3$\,\pcc, and $B_0=3$\,\muG.
Note the split log--linear $x$-axis.
\label{fig:Prof_M_En}}
\end{figure}

The momentum, $\PhiPmc(x)$, and energy, $\PhiEmc(x)$, fluxes from
all particles are calculated  directly in the \mc\ simulation for
the $i$-th iteration. To these particle fluxes we add the wave
components $\PwI(x)$ and $\FwI(x)$ and check to see if the total
momentum and energy fluxes (i.e., equations~\ref{momentunFlux1} and
\ref{momentunEnergy1}) are conserved  to within some limit at all
$x$. If the fluxes are not conserved in the \hbox{$i$-th}
iteration, we use equation~(\ref{momentunFlux1}) in the form
\begin{equation}
\rho_0 u_0\, [u^{(i+1)}(x) - u^{(i)}(x)] + \PhiPmc(x) +
\PwI(x)=\PhiPz \ ,
\end{equation}
to predict the shock speed profile, $u^{(i+1)}(x)$, for the
\hbox{$i$-th$+1$} iteration.

When \rel\ particles are produced and/or high-energy particles
escape  at an upstream FEB, the overall compression, $\Rtot$, will
increase above $\rRH$  and $\Rtot$ must be found by
iteration simultaneously with $u(x)$ and $\Vscat(x)$. A consistent
solution, within some statistical uncertainty, will conserve
momentum and energy fluxes at all $x$, including a match to the
escaping energy flux, $\Qesc$ in equation~(\ref{momentunEnergy1}).
Despite the complexity of this system, with several processes all
coupled nonlinearly, we are able to obtain unique modified shock
solutions covering a wide dynamic range, with \SC\ injection, MFA,
and an overall compression ratio consistent with particle escape.

\begin{figure}
\epsscale{1.1}
\plotone{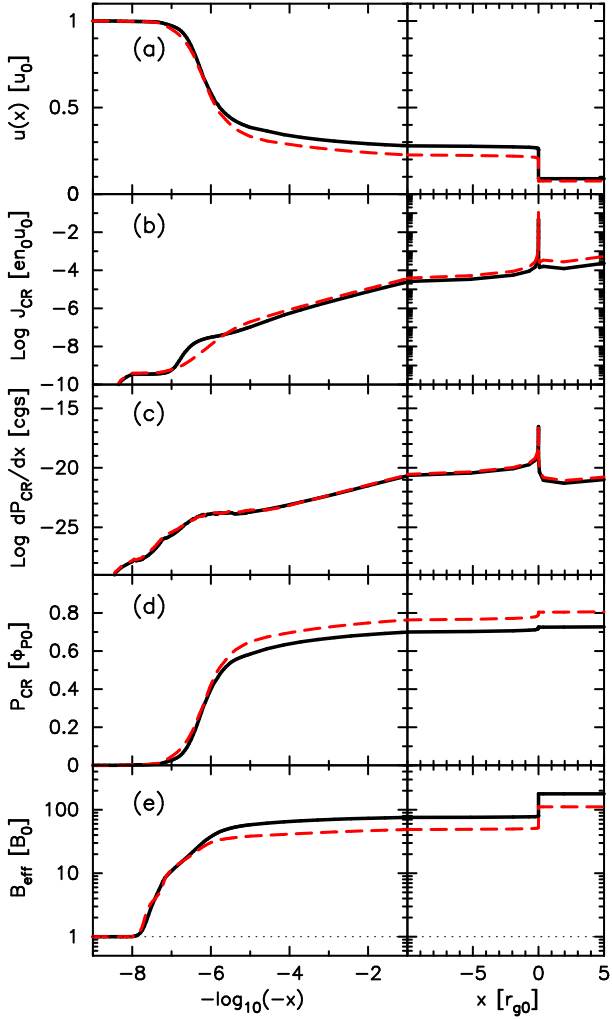}            % Fig 2
\caption{The top panels (a) show the bulk flow speed
as in Figure~\ref{fig:Prof_M_En}, panels (b) show the CR current, panels (c) show $d\Pcr/dx$ in cgs units, panels (d) show the CR pressure, and the bottom panels (e) show
the effective magnetic field derived from
equation~(\ref{eq:Beff}).
The dashed (red) curves include cascading (Model \aafive)
while the solid (black) curves (Model \aafour) do not.
Both models have $u_0=5000$\,\kmps, $n_0=0.3$\,\pcc, and $B_0=3$\,\muG.
\label{fig:Jcr_dPcr}}
\end{figure}

\begin{figure}   % Fig 3
\epsscale{1.1}
\plotone{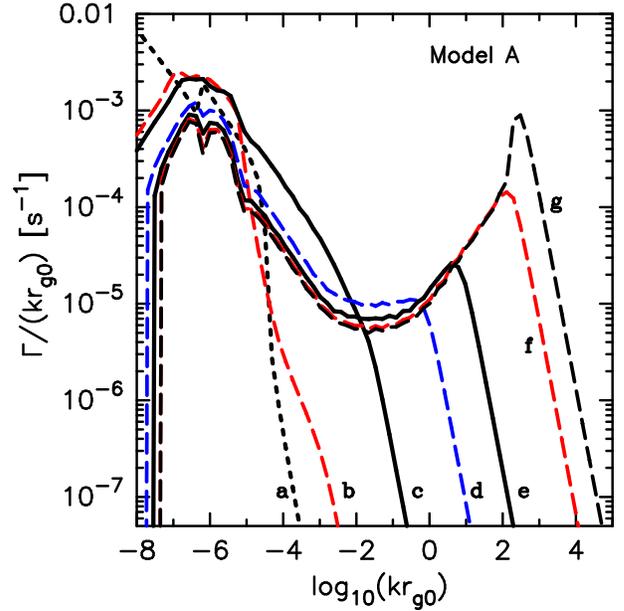} 
\caption{The instability growth rate at different positions in the shock precursor for our \NL\ Model \aafour. We plot  $\Gamma/(k \rgz)$
versus $k \rgz$  to reduce the scale spread and the labels
indicate the $x$-positions where the curves are calculated: (a) $10^8$, (b) $10^7$,  (c) $10^6$, (d) $1.9\xx{3}$, (e) $5.2$,
(f) $1.9\xx{-3}$ and (g) $10^{-4}$, all in units of $-\rgz$.
Compare this with Figure~1 in \citet{boe11}.
\label{fig:GamTau}}
\end{figure}

\begin{figure}
\epsscale{1.0}
\plotone{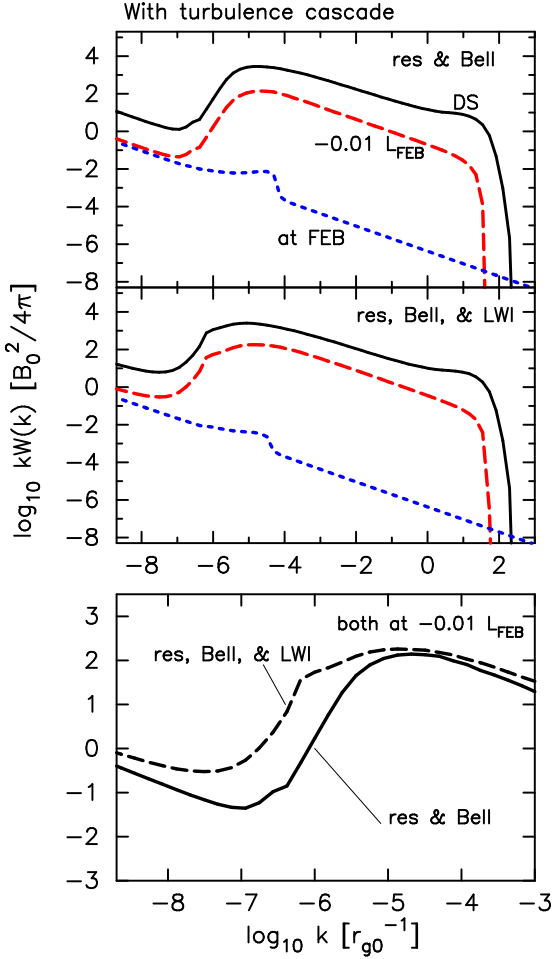}            % Fig 4
\caption{Turbulence spectra for the example shown in
Figure~\ref{fig:Jcr_dPcr} with turbulence cascading. In the top two panels, the solid (black) curve is calculated downstream from the subshock, the dashed (red) curve is calculated in the shock precursor at $x=-0.01\,\Lfeb= -10^6\rgz$, and the dotted (blue) curve is calculated at the FEB.  In the top panel (Model \aathree), the LWI is not included while in the middle panel (Model \aafour) it is. In all cases, the resonant instability is included. In the bottom panel, the cases with and without the LWI are compared  at $x=-0.01\,\Lfeb$.
\label{fig:Turb_yes_cas}}
\end{figure}

\begin{figure}
\epsscale{1.0}
\plotone{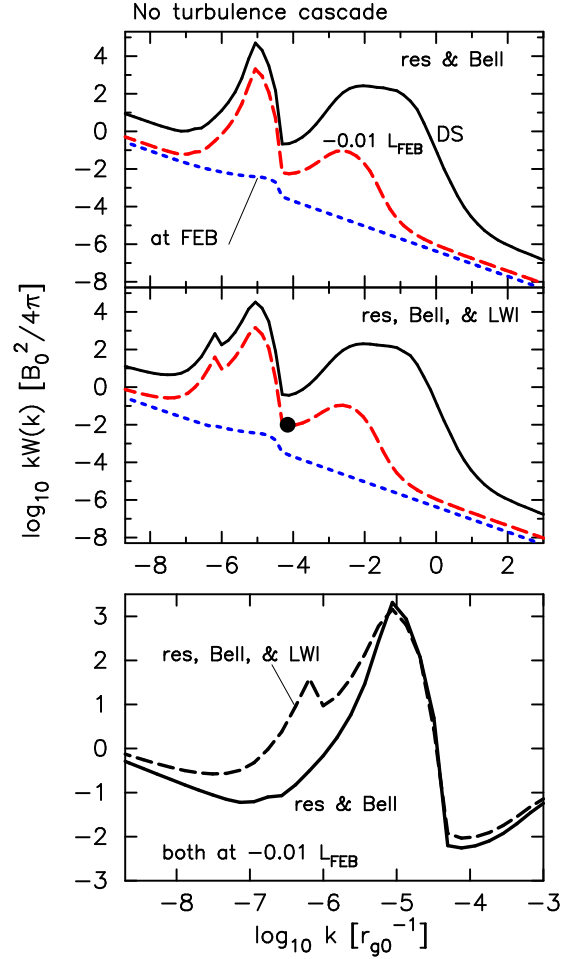}            % Fig 5
\caption{Turbulence spectra for the example shown in
Figure~\ref{fig:Jcr_dPcr} without turbulence cascading. Other than cascading, all aspects of the figure are
similar to Figure~\ref{fig:Turb_yes_cas}. The model without the LWI is Model \aatwo, while that with the LWI is Model \aafour.
The turbulence surrounding the solid dot in the middle panel corresponds roughly to the diffusion coefficient at the solid dot in the bottom panel of Figure~\ref{fig:fp_mfp_UpS}.
\label{fig:Turb_no_cas}}
\end{figure}

\begin{figure}
\epsscale{1.1}
\plotone{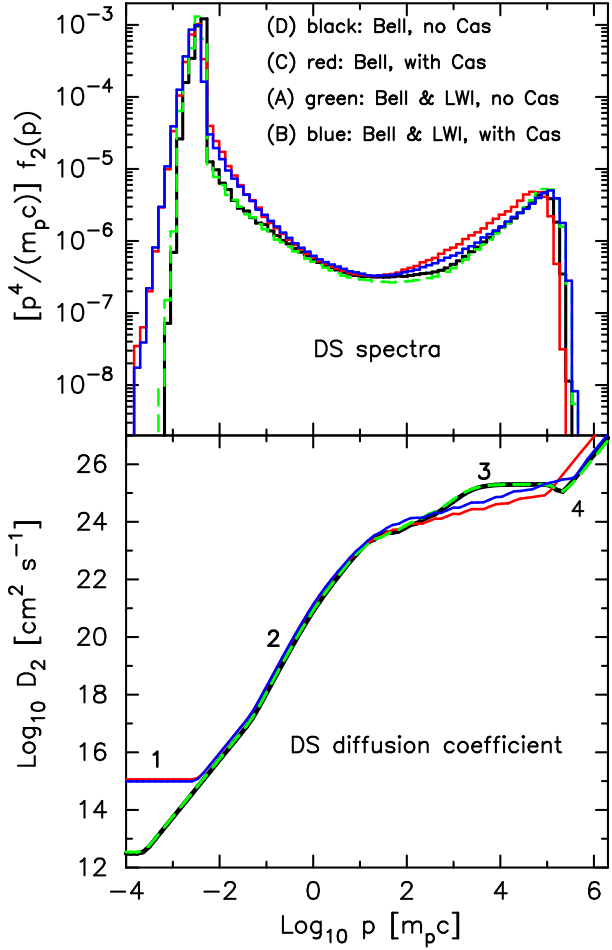}            % Fig 6
\caption{Phase-space distributions, $[p^4/(m_pc)] f_2(p)$, and diffusion coefficients  for the four examples shown in
Figures~\ref{fig:Turb_yes_cas} and \ref{fig:Turb_no_cas} calculated downstream from the subshock. The color coding is the same for each panel and the labeled regions in the bottom panel are discussed in the text. Note that the resonant instability is active in all examples. Here and elsewhere the particle distribution is anisotropic and calculated in the shock frame and the plotted $f_2(p)$ is  averaged over all angles and  normalized such that $n(x)=4 \pi \int_0^\infty f(x,p)p^2dp =n(x)$ is the local number density.
The models are: black (\aatwo), red (\aathree), green (\aafour), and blue (\aafive). \label{fig:fp_mfp_DS}}
\end{figure}

\begin{figure}
\epsscale{1.1}
\plotone{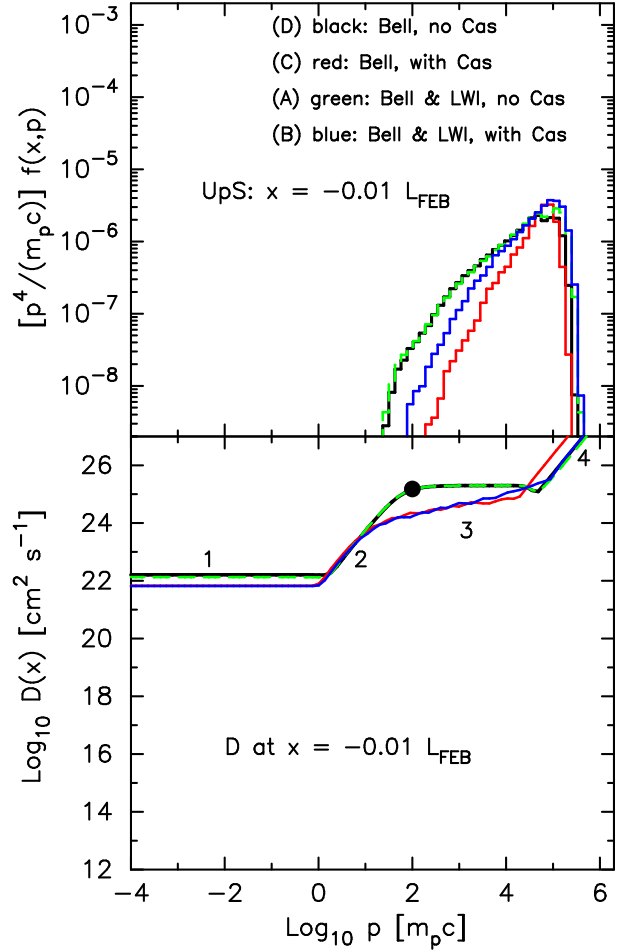}            % Fig 7
\caption{Phase-space distributions, $[p^4/(m_pc)] f(x,p)$, and diffusion coefficients  for the four examples shown in
Figures~\ref{fig:Turb_yes_cas} and \ref{fig:Turb_no_cas} calculated in the shock precursor at $x = -0.01\,\Lfeb$. The color coding, labels, and normalization of $f(x,p)$ are the same as in
Figure~\ref{fig:fp_mfp_DS} and the resonant instability is active for all examples. The models are: black (\aatwo),
red (\aathree), green (\aafour), and blue (\aafive). Note that the cold incoming beam is present in the precursor but not shown in the upper panel. The solid dot in the bottom panel corresponds to the solid dot in the middle panel of
Figure~\ref{fig:Turb_no_cas}
\label{fig:fp_mfp_UpS}}
\end{figure}

\begin{figure}
\epsscale{1.1}
\plotone{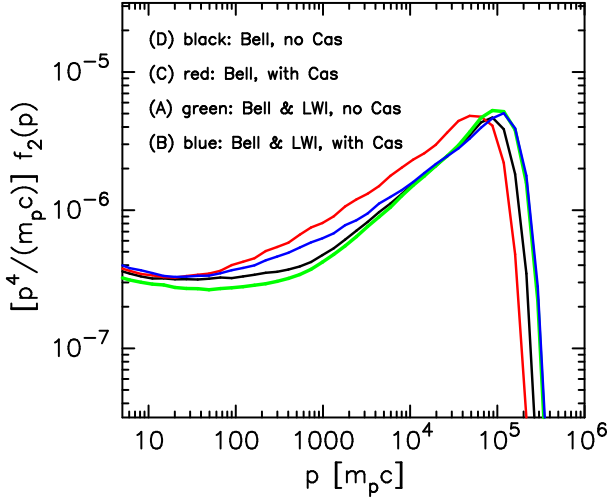}            % Fig 8
\caption{High momentum portion  of the downstream phase-space
distributions, $[p^4/(m_pc)] f_2(p)$, shown in
Figure~\ref{fig:fp_mfp_DS}. The resonant instability is active for
all examples. The models are: black (\aatwo), red (\aathree), green
(\aafour), and blue (\aafive). \label{fig:fp_pmax}}
\end{figure}

\begin{figure}
\epsscale{1.15}
\plotone{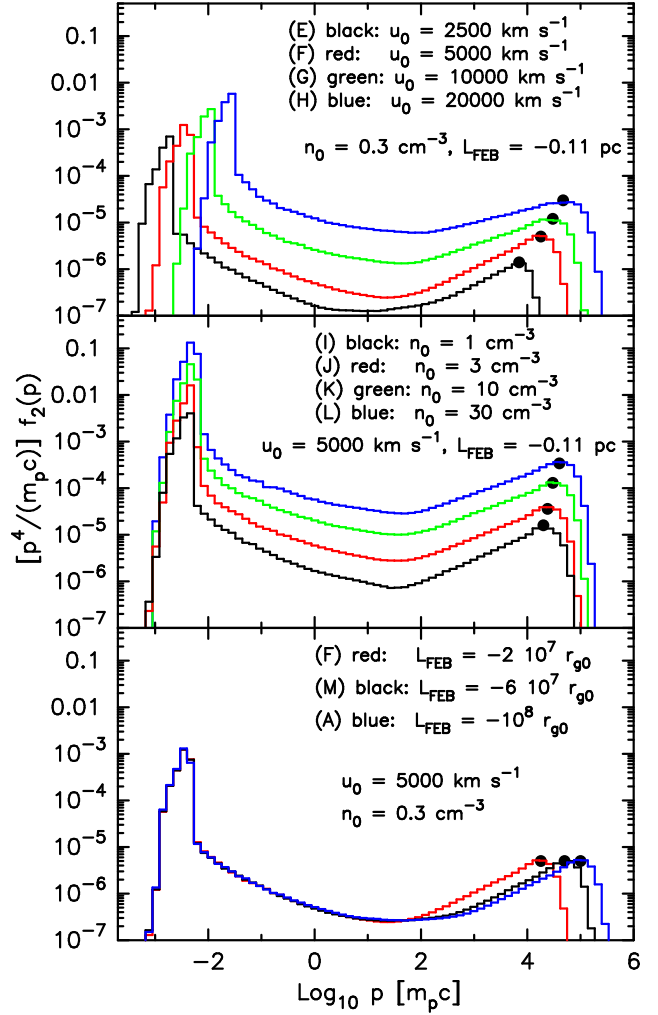}            % Fig 9
\caption{Downstream, shock frame proton  distribution functions
calculated without cascading for the models as indicated. In the top panel, the shock speed,
$u_0$, is varied, in the middle panel the ambient density, $n_0$, is
varied, and in the bottom panel, the FEB, $\Lfeb$, is varied.
Parameters held constant are shown in each panel.  In all cases,
$T_0=10^4$\,K and $B_0 = 3$\,\muG. The solid dots indicate where
$p^4 f_2(p)$ is a maximum and defines $\pmax$. \label{fig:pdf_vary}}
\end{figure}

\begin{figure}
\epsscale{1.1}                   % Fig 10
\plotone{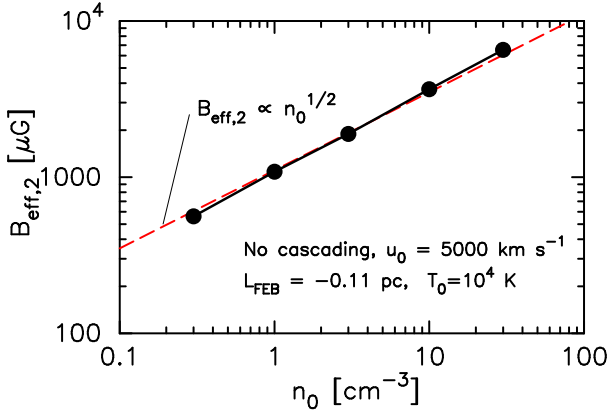} 
\caption{The downstream amplified field, $\BeffDS$, versus ambient density. The models running left to right are \aaseven, \aasix, \aaeight, \aanine, and \aaten.
%
%In the bottom panel the models running left to right are %\bbthree, \aaeleven, \aaseven, \aaone, \aasixteen, and %\aaseventeen. 
The dashed (red) line shows the function as indicated.
\label{fig:Bscale}}
\end{figure}

\begin{figure}
\epsscale{1.1}                   % Fig 11
\plotone{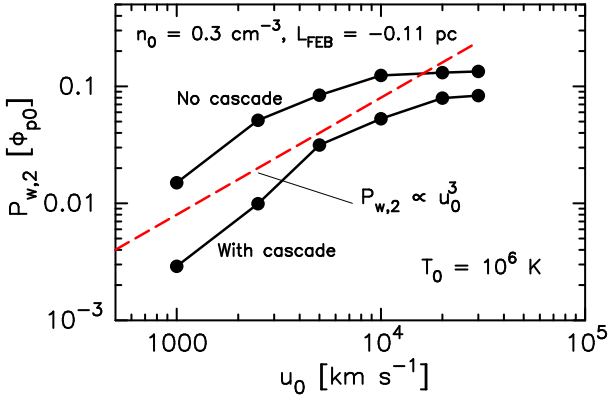} 
\caption{Downstream pressure in turbulence versus shock speed for the models (running left to right) \bbthree, \bbseven, \bbfive, \bbone, \bbnine, and \bbeleven\ without turbulence cascade and models \bbfour, \bbeight, \bbsix, \bbtwo, \bbten, and \bbtwelve\ with cascade. The dashed (red) line approximates the behavior for low shock speeds.
\label{fig:waveF}}
\end{figure}

\begin{figure}
\epsscale{1.1}                   % Fig 12
\plotone{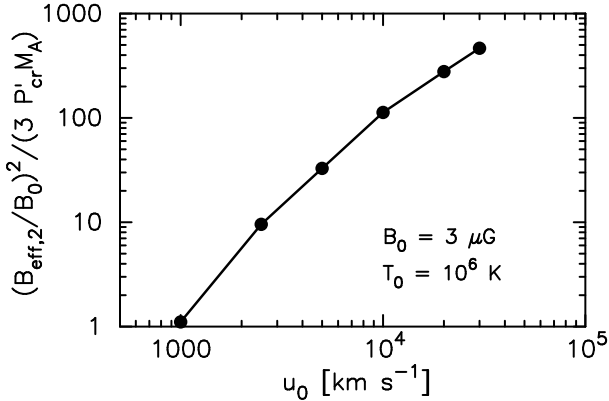} 
\caption{Comparison of our models (from left to right)
\bbthree, \bbseven, \bbfive, \bbone, \bbnine, and \bbeleven\
with equation~(2) in \citet{CS2014}. Note that $\Pcr'=\Pcr/(\rho_0 u_0^2)$ plotted in the $y$-axis is the normalized CR pressure.
\label{fig:Cap}}
\end{figure}

\begin{figure}
\epsscale{1.1}                   % Fig 13
\plotone{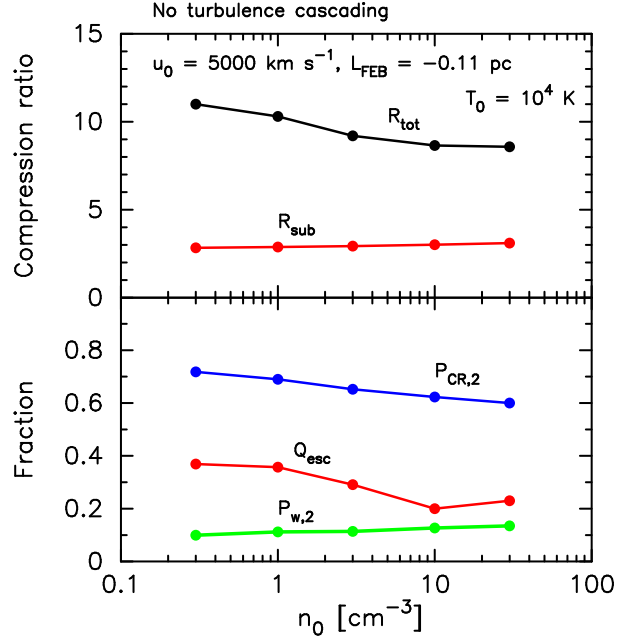} 
\caption{The top panel shows the total and subshock compression ratios and the bottom panel shows the acceleration efficiency as given by the DS pressure in trapped CRs $\PcrDS$, the DS pressure in turbulence $\PwDS$ (both as fractions of $\PhiPz$),
and the fraction of upstream energy flux escaping at the upstream FEB
$\Qesc$ versus ambient density for the models (from left to right) \aaseven, \aasix,
\aaeight, \aanine, and \aaten.
\label{fig:Eff_scale_n0}}
\end{figure}

\begin{figure}
\epsscale{1.1}                   % Fig 14
\plotone{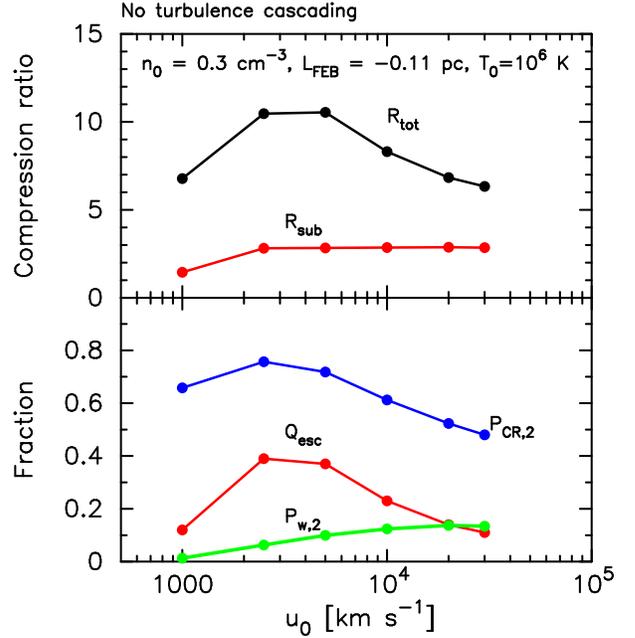} 
\caption{The top panel shows the total and subshock compression ratios as a function of shock speed for fixed $n_0$ and physical distance to the FEB. In the bottom panel various quantities are shown as labeled. The CR and wave pressures are calculated downstream from the
subshock as fractions of $\PhiPz$ and $\Qesc$ is given as the fraction of $\PhiEz$. The models, running left to right, are \bbthree, \bbseven, \bbfive, \bbone, \bbnine, and \bbeleven.
\label{fig:Eff_scale_u0}}
\end{figure}

\begin{figure}
\epsscale{1.1}                   % Fig 15
\plotone{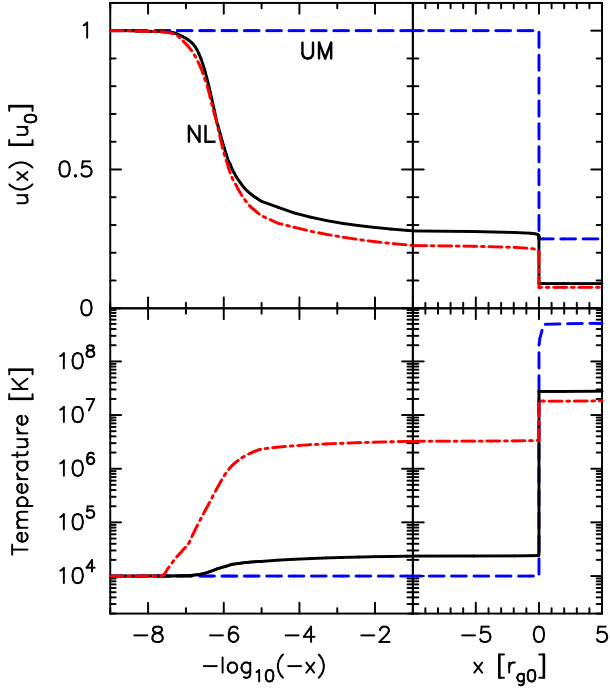} 
\caption{The top panel compares the shock structure for an unmodified shock (model \aafourUM, dashed blue curve), a modified shock without cascading (model \aafour, solid black curve), and a modified shock with cascading (model \aafive, dot-dashed red curve)
In the bottom panel, the temperature structure is shown using the same line notation.
\label{fig:Temp}}
\end{figure}

\begin{figure}
\epsscale{1.1}                   % Fig 16
\plotone{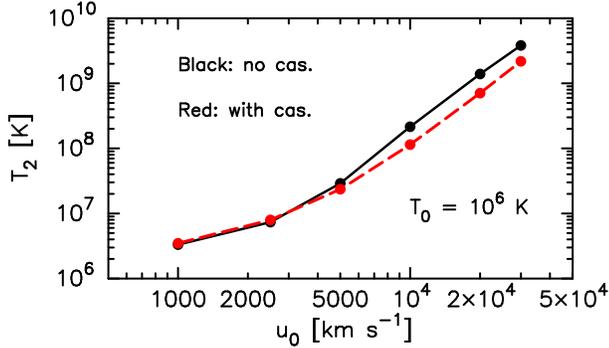} 
\caption{Downstream temperatures with (dashed curve) and
without (solid curve) cascading. The models without cascading, running left to right, are \bbthree, \bbseven, \bbfive, \bbone, \bbnine, and \bbeleven.
Those with cascading are \bbfour, \bbeight, \bbsix, \bbtwo, \bbten, and \bbtwelve.
\label{fig:T2_vs_u0}}
\end{figure}

\begin{figure}
\epsscale{1.1}                   % Fig 17
\plotone{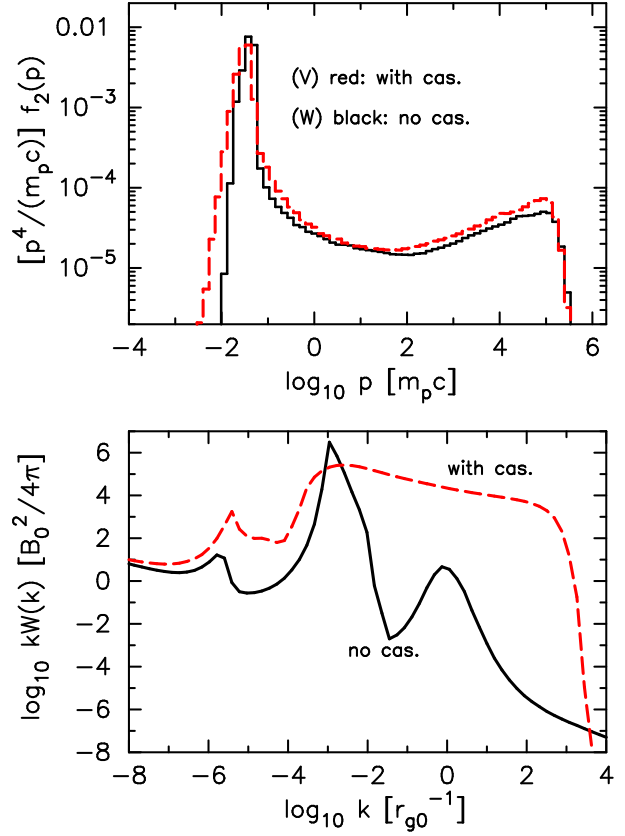} 
\caption{The top panel shows the downstream phase-space distributions and the bottom panel shows the downstream turbulence for models \bbeleven\ and \bbtwelve. These models have $u_0=3\xx{4}$\,\kmps\ and $\Lfeb\simeq 0.11$\,pc.
\label{fig:fp_30000}}
\end{figure}

\begin{figure}
\epsscale{1.1}                   % Fig 18
\plotone{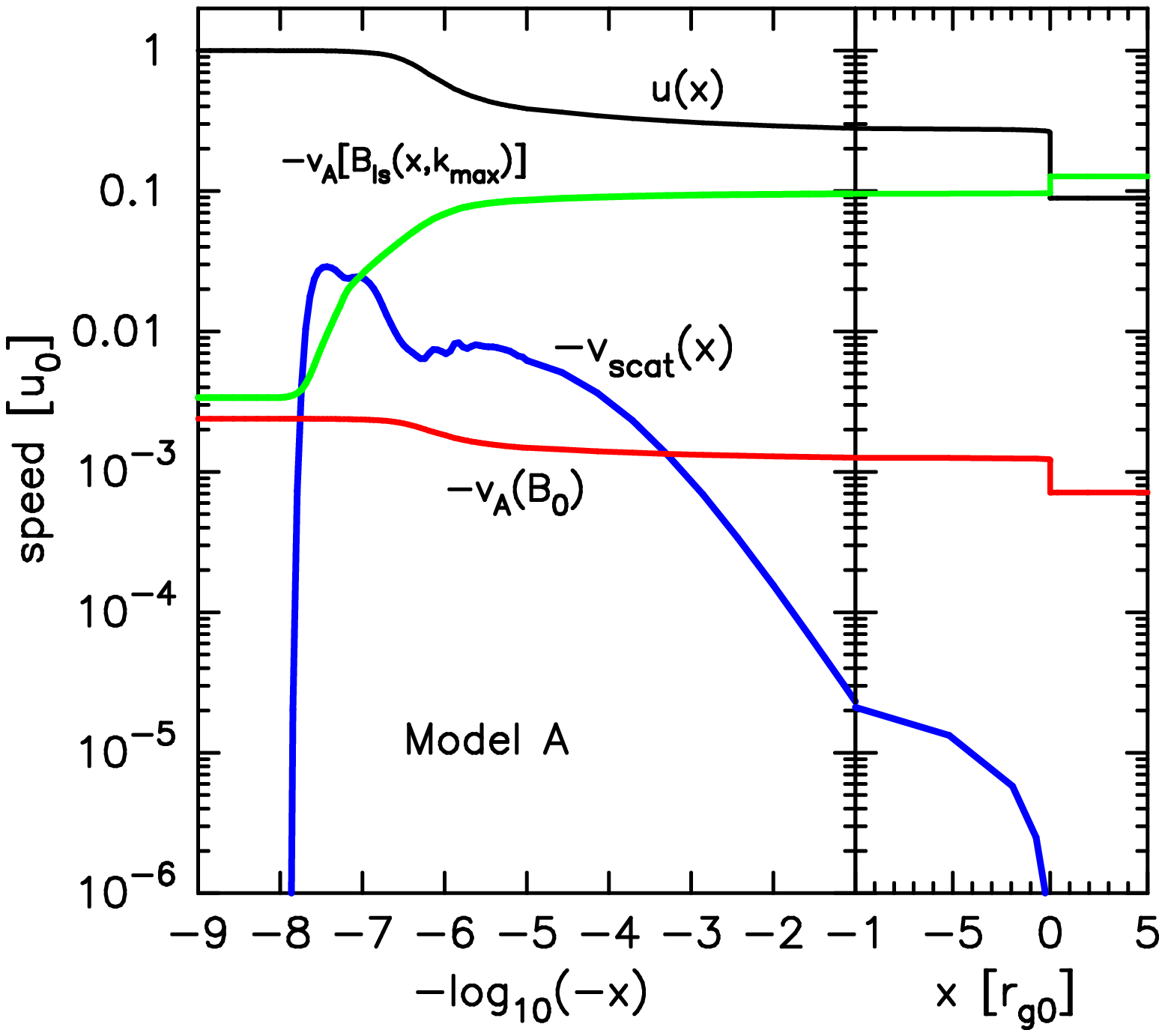} 
\caption{The scattering center velocity  $\Vscat(x)$  (blue line)
derived from our Model \aafour\ for a shock velocity
$u_0=5\xx{3}$\,\kmps,
far upstream density $n_{0}=0.3$\,\pcc, far upstream magnetic field
$B_0 = 3$\,\muG, and
$\Lfeb\simeq 0.56$\,pc.  We also show the bulk flow velocity $u(x)$
(black line) and the \alf\ velocities calculated with the amplified magnetic field $\Bls(x,\kmax)$ (green line) and the initial field $B_0$
(red line). Note that $\Vscat$ and the \alf\ speeds are directed upwards so we plot the negative values.\label{fig:vscat}}
\end{figure}

\section{Results} \label{sec:results}
In our solutions,  the bulk flow speed, $u(x)$, overall compression ratio, $\Rtot$, and CR induced magnetic turbulence are calculated such that
the momentum and energy fluxes are conserved across the shock, as
shown by the solid (black) curves in
Figure~\ref{fig:Prof_M_En} (model \aafour\ in
Table~\ref{tab:param}).
These fluxes include the magnetic field contributions and account
for the escaping energy flux at the FEB. The drop in energy flux seen in the solid curve at $x\sim -10^8\,\rgz$ in the bottom left panel of
Figure~\ref{fig:Prof_M_En} is a direct measure of the energy flux escaping at the FEB.

The dashed (red) curves in
Figure~\ref{fig:Prof_M_En} (model \aafourUM\ in Table~\ref{tab:param}) show the shock structure and fluxes for the same input parameters without shock smoothing, adjustment of $\Rtot$, or MFA.
In the UM case, $\Rtot = \rRH \simeq 4$, versus $\Rtot \simeq 11.3$ in the NL case, and the momentum and energy fluxes are $\sim 100$ times above the conserved values throughout much of the shock. These quantitative  results, of course, depend on the efficiency of DSA which, in turn, depends on the details of our model, i.e., on the injection scheme, the MFA description, and the calculation of the scattering \mfp\ from the turbulence.  Nevertheless, it is essential to note that any consistent description of efficient DSA with a diffusion coefficient that is an increasing function of momentum must produce a shock structure similar to the solid (black) curves shown in
Figure~\ref{fig:Prof_M_En}.

In the \SC\ solution, the shock structure develops a distinct
subshock with compression ratio $\Rsub < \Rtot$, as indicated by
$u_1$ in the top panel of Figure~\ref{fig:Prof_M_En}. The
subshock is responsible  for most of the heating of the ambient
material and in this case, $\Rsub = u_1/u_2 \simeq 2.85$. In order
to have a consistent solution with efficient DSA, the plasma heating
must be reduced compared to the UM case, as reflected by $\Rsub < \Rtot$.

\subsection{Magnetic Field Fluctuations Spectra} \label{sec:Flux}
Our model gives a thorough accounting of the
magnetic fluctuation amplification produced
simultaneously by the three CR-current
instabilities as discussed in \S\ref{sec:Grow}.
In Figure~\ref{fig:Jcr_dPcr} we show the converged flow speed
profile  along with the CR current $\Jcr$, CR pressure gradient
$d\Pcr/dx$, CR pressure $\Pcr$, and the effective magnetic field $\Beff$
that results
from MFA. Note that $\Beff$ is obtained from equation~(\ref{eq:Beff2}) and does not include the homogeneous part of the ISM field.
The dashed (red) curves (Model \aafive\ in Table~\ref{tab:param}) include cascading while the
solid (black) curves (Model \aafour) do not.
From the (d) and (e) panels it can be seen that cascading reduces the large-scale magnetic field by about a factor of two for $x \gsim -10^{6}\,\rgz$ and causes the CR pressure to increase by $\sim 10\%$ over the same range.
In both cases, the shock structure (panel a) adjusts so momentum and energy are conserved through the shock.

In Figure~\ref{fig:GamTau} we show
$\Gamma(x,k)/( \rgz)$ vs. $k \rgz$ as
calculated from equation~(\ref{SolveDispersMedium}) for
Model~\aafour.
Here, $\Gamma(x,k) = 2\, \mathrm{Im}[\omega(x,k)]$ is the growth
rate of magnetic energy of the modes.
The curves are calculated at different positions in the precursor
going from the FEB at $x=-10^8\rgz$ (a)  to just upstream of
the subshock position at $x=10^{-4}\rgz$ (g).
It is instructive to compare this figure with Figure~1 in
\citet{boe11} where the growth rates for the three instabilities are
plotted separately for a fixed set of parameters.
In the \mc\ code, the combined turbulence growth
rate is determined \SCly\ from the three instabilities at each position in the shock precursor and the instantaneous growth rate varies widely as a function of position and wave number.

In addition to modifying the MFA, the efficiency of turbulence cascade
through $k$-space strongly influences the spectrum of mesoscopic magnetic fluctuations.
Unfortunately, the nature of turbulence cascade is not well understood. Local turbulence cascade parallel
to the mean field can be suppressed in MHD turbulence
\citep[e.g.,][]{goldr97,Biskamp2003,Sahraouiea06}.
We assume this to justify our models without cascade where we set $\Pi(x,k)=0$ in equation~(\ref{eq:EnBal}).
Both the Bell and long-wavelength  instabilities  have maximum  growth rates along the local mean magnetic field \citep[e.g.,][]{bbmo13}.
We leave the more difficult issue of  anisotropic cascading to future work.

In Figures~\ref{fig:Turb_yes_cas} and \ref{fig:Turb_no_cas}
we show the turbulence, with and without cascading, calculated upstream at the FEB (dotted, blue curves), in the precursor at 1\% of the distance to the FEB (dashed, red curves), and  downstream from the shock
(solid, black curves).\footnote{Except for the small bump produced by escaping CRs, the dotted (blue) curves at the FEB are essentially the background turbulence assumed in the simulation. The far upstream turbulence may well be modified by the escaping CR flux \citep[e.g.,][]{SB2014} but we do not consider that here.}
As expected, cascading produces large differences in the  turbulence at short wavelengths but the longest wavelengths are much less affected.
Also shown in
Figures~\ref{fig:Turb_yes_cas} and \ref{fig:Turb_no_cas} is the effect of the LWI. The top panels show the turbulence without the
LWI ($N_B=0$), the middle panels show the case where resonant, Bell, and the LWI are combined, and the bottom panels show a long-wavelength
comparison in the precursor at $x = -0.01\,\Lfeb$.
The comparison at $x = -0.01\,\Lfeb$ in the bottom panels
shows that the LWI
broadens the spectral peak and shifts it toward larger scales.
The LWI enhances the longest wavelength turbulence by at least an order of magnitude with or without cascading.

In the top panel of Figure~\ref{fig:fp_mfp_DS} we show the shock
frame, downstream proton phase-space distributions, multiplied by
$p^4/(m_pc)$, for the four cases shown in
Figures~\ref{fig:Turb_yes_cas} and \ref{fig:Turb_no_cas}.
The bottom panel shows the downstream diffusion
coefficient, $D_2$, for
these cases where $D_2$, or $\lambda(x,p)$,  is determined from
equation~(\ref{eq:EffMFP}).
Figure~\ref{fig:fp_mfp_UpS} is a similar plot calculated in the shock precursor at $x=-0.01\,\Lfeb = -10^6\rgz$.
The regions indicated by numbers in
Figures~\ref{fig:fp_mfp_DS} and \ref{fig:fp_mfp_UpS} have particular characteristics. The low momentum region 1 is where vortex convection dominates and $D_2$ is independent
of $p$ (equation~\ref{eq:Lvor}).
For the upstream position shown in Figure~\ref{fig:fp_mfp_UpS}, low-energy accelerated particles do not reach this position so there is no CR pressure gradient or current to produce turbulence on short scales.
The highest energy region 4, is dominated by short-scale fluctuations
(equation~\ref{MFP_s}) and $D_2 \propto p^2$.
The intermediate range 2--3 is where quasi-resonant  processes
dominate and the $p$ dependence of $D_2$ depends on the interplay of
$\Bls$, $k$, and $W$ in equation~(\ref{eq:Lres}). However, since
$\Bls$ depends on $W$ and $k$ through equation~(\ref{B_ls}) there is
no simple one-to-one correspondence between the turbulence shown in
Figures~\ref{fig:Turb_yes_cas} and \ref{fig:Turb_no_cas} and $D_2$.

Nevertheless features such as the flattening of $D_2$ between
regions 2 and 3 can be understood in general terms. In going from low to high $p$, the resonant $k$ decreases causing $\Bls$ to decrease
(equation~\ref{B_ls}). This, combined with the increase (or slow decrease) in $kW$ as $k$ decreases causes $\lambda(x,p)$
(equation~\ref{eq:EffMFP}) to increase slowly.
Here, in region 3, $D_2$ increases less rapidly then $p$, i.e., slower than the Bohm limit. In contrast, in region 2,  as $p$ increases, $k$ decreases, $kW$ flattens out and $D_2$ increases faster than $p^2$. The transition
in $D_2$ from 2 to 3 indicated with a solid dot in the bottom panel of
Figure~\ref{fig:fp_mfp_UpS} corresponds roughly to the turbulence at
the position indicated by the solid dot in the middle panel of
Figure~\ref{fig:Turb_no_cas}.

While the  magnetic fluctuation spectra with  and
without cascading are very different in the short-scale regime, as shown in
Figures~\ref{fig:Turb_yes_cas} and \ref{fig:Turb_no_cas}, the
corresponding DS particle spectra, shown in
Figure.~\ref{fig:fp_mfp_DS}, are
quite similar.
There is, however, a clear increase in $\pmax$ when the LWI operates with the Bell and resonant instabilities. This is shown in
Figure~\ref{fig:fp_pmax}, where $\pmax$ increases by about a factor of two when the LWI is included.
This coupling between the short-wavelength modes from  Bell's
instability (much shorter than the CR gyroradius) and the
long-wavelength modes from the LWI (much longer than the CR
gyroradius) highlights the need for simulations to cover a  wide
dynamic range in order to capture this essential physics.

\subsection{Particle spectra}
In Figure~\ref{fig:pdf_vary}  we show downstream
particle spectra where the resonant, Bell, and long-wavelength  instabilities are active
for shocks with varying speed (top panel), varying
ambient density (middle panel), and varying $\Lfeb$ (bottom panel.
For all plots there is no cascading, $T_0=10^4$\,K,
$B_0=3$\,\muG, the ISM turbulence is such that $\Bkolm=3$\,\muG, and for the top two panels the FEB is at the same physical distance from the subshock, i.e., $\Lfeb \simeq -0.11$\,pc.

In contrast to the top panel in
Figure~\ref{fig:fp_mfp_DS}, where the downstream distribution functions vary little with changes in the turbulence generation and cascading, the spectra in Figure~\ref{fig:pdf_vary}
vary importantly with $u_0$, $n_0$, and $\Lfeb$. For a given physical distance to the FEB (top panel), the maximum CR momentum, $\pmax$, increases with $u_0$.
We indicate the trend in $\pmax$ with a solid dot placed at the maximum in $p^4 f(p)$.
This trend results from the fact that the upstream diffusion length  scales as $1/u_0$ so a fixed
$\Lfeb$ increases in terms of particle gyroradii as $u_0$ increases.
The top panel also shows that the thermal peak and the minimum in the $p^4 f(p)$ distribution increase with $u_0$. It is significant that the minimum in $p^4 f(p)$ occurs well above $m_pc$ in all cases.

The middle and bottom panels of Figure~\ref{fig:pdf_vary} show that the normalization of $f_2(p$) scales as $n_0$ and
$\pmax$ scales both with $\Lfeb$ and $n_0$. From the three sets of simulations
in Figure~\ref{fig:pdf_vary} (and additional runs not shown for clarity),
we obtain the scaling relation
\begin{equation}\label{p_m}
\pmax \propto n^\delta_0 \,u_{0} \Lfeb
\end{equation}
or, assuming that $\Lfeb$ is some fraction of the shock radius $\Rsk$, one obtains
$\pmax \propto n^\delta_0 \, u_{0} \Rsk$.
Notably, we find a  rather weak dependence of $\pmax$ on $n_{0}$
 of $\delta \sim$ 0.25 (middle panel of Figure~\ref{fig:pdf_vary}).
This result, for a quantity critical  for all shock applications, is in contrast with the scaling
expected if one assumes $D(\pmax)/u_{0} \propto \Rsk$ and
that $D(\pmax)$ depends on the proton gyroradius $\rg(\pmax)$ in
the effective, downstream, amplified magnetic field, $\BeffDS$.
Since MFA depends on the ram pressure,  it has been suggested that  $\BeffDS \propto \sqrt{n_{0}}\, u_{0}$
\citep[see, e.g.,][]{pzs10,SchureBell2013}.
In this case, $\pmax$ would scale as
$\sqrt{n_{0}}\, u_{0}^2\, \Lfeb$, a stronger dependence on both the upstream density  and the shock velocity then we
find with our \SC\ simulations.

In regards to the postshock turbulent magnetic field, $\BeffDS$, we find the following dependence on the far upstream density and shock velocity (see Figures~\ref{fig:Bscale} and \ref{fig:Cap})
\begin{equation} \label{eq:Bscale}
\BeffDS \propto \sqrt{n_{0}}u_{0}^\theta
\ .
\end{equation}
At  $u_0 \gsim 5,000$\,\kmps, the efficiency of MFA,
defined as $\PwDS/\FpxZ$,
saturates at roughly  10-15\% of the far upstream ram
pressure (see Figure~\ref{fig:waveF}).
This implies $\PwDS \propto u_0^2$, or
$\BeffDS^2 \propto u_0^2 n_0$,
corresponding to $\theta \sim 1$.
At lower shock velocities, we find
$\PwDS/\FpxZ \propto u_0$ (i.e., $\PwDS \propto u_0^3$),
and therefore $\BeffDS^2/n_0 \propto u_0^3$,
i.e., $\theta \sim $ 1.5.
At shock
velocities below $\sim 1000$\,\kmps,  the efficiency of MFA drops to about a percent of the ram pressure.

In Figure~\ref{fig:Cap} we compare our results with the scaling relation determined in \citet{CS2014} using hybrid simulations. Their
equation~(2) (with our notation) is
%
%\begin{equation} \label{eq:CapEq}
%\left < \frac{\Btot}{B_0} \right >^2_\mathrm{sh} \simeq
%3 \frac{\PcrDS}{\PhiPz} M_A
%\ ,
%\end{equation}
%
%
\begin{equation} \label{eq:CapEq}
\left < \frac{\BeffDS}{B_0} \right >^2_\mathrm{sh} \simeq
3 \frac{\PcrDS}{\PhiPz} M_A
\ ,
\end{equation}
and it's clear from Figure~\ref{fig:Cap} that, while the \mc\ result matches equation~(\ref{eq:CapEq})
at $u_0=1000$\,\kmps\ (i.e., $M_A = 84$), it has a very different scaling at higher shock speeds and higher \alf\ Mach numbers.
It is important to understand the reasons for this difference, which must stem from the different  assumptions and parameters for the two simulations, since any modeling of a real object requires such scalings.
Our \alf\ Mach numbers run from about 84 to 2500, whereas the hybrid results are for $M_A \lsim 100$.
This is significant because MFA in shocks with \alf\ Mach numbers below about 30 is dominated by the resonant CR instability
\citep[e.g.,][]{ab09,CS2014}.
In our high Mach number results, non-resonant instabilities dominate.
Apart from different magnetizations, there are different assumptions made for the magnetic fluctuation spectra of the incoming plasma.
In the \mc\ modeling, the initial spectrum of magnetic fluctuations is Kolmogorov, as determined by
equation~(\ref{Bibint}), whereas the initial fluctuations in the hybrid simulations are determined by numerical noise.

It is significant that the dependence of the
post-shock turbulent magnetic field can be tested with SNR
observations. The compilation by \citet[][]{vink12}
(see his Figure~39) can be understood if the scaling
$\PwDS \propto u_0^3$
(i.e., $\BeffDS^2 \propto u_0^3 n_0$) holds at shock
velocities below $\sim 10^4$\,\kmps\ and changes
to $\PwDS \propto u_0^2$ (i.e., $\BeffDS^2 \propto u_0^2 n_0$) at
higher shock velocities.
We find that this scaling holds for both upstream
temperatures $T_0= 10^4$ and $10^6$\,K, indicating  that it is not
a sonic Mach number effect, at least for fairly large $M_S$.
We also find that it is only weakly dependent on the  far upstream
mean field $B_0$, indicating only a weak \alf\ Mach number dependence. In addition, the scaling doesn't depend strongly on $\Lfeb$.

Apart from the magnetic field--shock velocity scalings, high spatial resolution {\sl Chandra} X-ray observations of Tycho's SNR  reported by \citet{Eriksen11} have revealed coherent synchrotron structures which may be related to amplified magnetic
fields \citep[e.g.,][]{Tycho_stripes11,msd12}. The presence of
long-wavelength peaks in magnetic turbulence spectra, which are apparent in
Figures~\ref{fig:Turb_no_cas} and \ref{fig:fp_30000}
(bottom panel), may be tested with high resolution synchrotron images. We shall discuss the modeling of synchrotron images elsewhere.

Other scalings include the acceleration  efficiency, as indicated by
the escaping CR energy flux $\Qesc$,\footnote{The escaping CR energy flux in Figures~\ref{fig:Eff_scale_n0}
and \ref{fig:Eff_scale_u0} and in Table 1 are defined in the shock rest frame. The escaping energy flux in the far upstream (i.e., ISM for SNRs) frame is also meaningful.
These two fluxes are very close for shocks with velocities
below $\sim 10^4$\,\kmps,  but the difference can become important for higher speeds. At $u_0 = 3\xx{4}$\,\kmps, the escaping flux in the ISM rest frame is $\sim$ 30\% higher than in the shock frame.}
the DS CR pressure $\PcrDS$, the DS wave
pressure $\PwDS$, and the shock compression ratios, $\Rtot$ and
$\Rsub$.
In {Figures~\ref{fig:Eff_scale_n0} and
\ref{fig:Eff_scale_u0} we show these scalings for shocks without cascading and for $\Lfeb \simeq -0.11$\,pc.
We find that increasing $n_0$
from 0.3 to 30\,\pcc\ (Figure~\ref{fig:Eff_scale_n0})
results in a decrease of $\Rtot$ from $\sim 11$ to $\sim 8.6$ with  a corresponding increase in $\Rsub$ from $\sim 2.8$ to $\sim 3.1$.
In Figure~\ref{fig:Eff_scale_u0} we see that as $u_0$ and the shock strength increase, $\Rtot$ and $\Qesc$ first increase and then decrease. Since $\Lfeb$ is set at a fixed physical distance for these examples, the size of the shock acceleration site
decreases with increasing $u_0$ resulting in a lower $\pmax$, as shown in the top panel of
Figure~\ref{fig:pdf_vary}. The decrease in $\pmax$ dominates the increase in shock strength causing $\Qesc$ to decrease when $u_0 \gsim 2500$\,\kmps.

\subsubsection{Cascading and Thermodynamic Properties}
In Figure~\ref{fig:Temp} we show  the effects of cascading
and the smooth shock structure on the background plasma temperature.
The dashed (blue) curves are for an unmodified shock (Model
\aafourUM), the solid (black) curves are for a modified shock
without cascading (Model \aafour), and the dot-dashed (red) curves
are for a modified shock with cascading (Model \aafive).
As in \citet{vbe09}, without cascading the dissipation term, $L$, in
equation~(\ref{eq:EnBal}) is set  to zero. When cascading is
included, we assume viscous dissipation such that $\Lwc = v_a k^2
k_d^{-1} W$ \citep[e.g.,][]{Vainshtein1993}.
The wavenumber, $k_d$, is identified with the inverse of the thermal
proton gyroradius: $k_d(x) = e\Bls(x,k_d)/[c\sqrt{m_pk_B T(x)}]$,
where $T(x)$ is the local gas temperature determined from the
heating induced by $L$, as described in \citet{vbe08}.

Without cascading, or in the UM shock, the  precursor temperature
remains within a factor of 3 of $T_0$ until a sharp increase occurs
at the subshock. In contrast, cascading heats the precursor
substantially so $T(x) \gsim 100 T_0$ well in front of the subshock which may produce observable consequences.
The downstream temperature is dramatically reduced in  the \NL\
cases compared to the UM shock and $T_2$ is slightly less with
cascading than without.

In Figure~\ref{fig:T2_vs_u0} we compare the  downstream
temperature for a different set of shocks with and without
turbulence cascading as a function of $u_0$. It's clear that $T_2$ is not very sensitive to cascading despite the large
effect on the precursor temperature, although the difference is
larger for larger $u_0$.
In the top panel of Figure~\ref{fig:fp_30000}, we show DS proton spectra for $u_0=3\xx{4}$\,\kmps. The fluxes at the
high-energy end of the spectra  are somewhat sensitive to the turbulent
cascading, while the maximum proton energy is not.
As we have emphasized earlier, \NL\ effects damp changes that might otherwise be expected from the large differences in the
self-generated turbulence (bottom panel). If the acceleration is as efficient as we show here, conservation considerations force the shock to adjust such that the particle distribution functions cannot vary substantially.
In contrast to the turbulence for the $u_0=5\xx{3}$\,\kmps\ shock shown in Figure~\ref{fig:Turb_yes_cas}, the spectrum with cascading for $u_0=3\xx{4}$\,\kmps\ shows a clear  long-wavelength peak (dashed red curve in the bottom panel of Figure~\ref{fig:fp_30000}).

\subsection{Scattering Center Velocity}
As described in Section~\ref{sec:ScatV}, we  calculate the
scattering center velocity, $\Vscat(x)$, from conservation
considerations without assuming any specific form for the
turbulence, in particular, without assuming \alf\ waves.
This macroscopic approach guarantees that the energy in the growing magnetic turbulence, which produces the CR scattering, is taken from the free energy of the anisotropic CR distribution.

In contrast, if the  scattering center
velocity is calculated in a microscopic, quasi-linear approach, one has to weight
the phase velocities of the modes with the anisotropy of their
wave-vector distributions.
In the simplest case of the resonant CR streaming
instability, the amplified modes are assumed to be \alf\ modes
with wave-vectors aligned against the CR pressure gradient in the
shock precursor. The situation is much more complicated for the dominant Bell and long wavelength nonresonant instabilities.
For these, no simple \alf-wave-like assumption is adequate.

In Figure~\ref{fig:vscat} we show $\Vscat(x)$ from Model \aafour\ along  with the
mean flow speed, $u(x)$, and the \alf\ speeds derived with the
upstream field $B_0$ and with the local amplified field
$\Bls(x,\kmax)$. It is seen that  $\Vscat(x)$ is everywhere well
below $u(x)$  so there is no sizeable CR spectral softening
due to a finite scattering center velocity.
The \mc\ $\Vscat(x)$ exceeds the \alf\ speed  calculated with $B_0$
in most of the precursor, but is well below the \alf\ speed
determined by the amplified $\Bls(x,\kmax)$ except in the far
upstream region near the FEB.
If $\Valf[\Bls(x,\kmax)]$ gives the scattering  center speed,
strong MFA implies significant softening of the CR spectrum since
the effective compression ratio for DSA will be less than $\Rtot$.
Such softening has been discussed by
\citet[][]{zp08a}, \citet[][]{blasi13}, and \citet[][]{kangea13}.

\citet[][]{mc12} have suggested that a large  scattering speed
resulting from MFA and rapid \alf\ waves in DSA with CR acceleration efficiencies $ \simeq 20$\%, could produce CR spectra steeper than $dN/dE \propto E^{-2}$.
If so, this might address the issue of the steep CR spectrum derived
from Fermi observations of Tycho's SNR \citep[see
also][]{Caprioli12,Slane2014}.
However, in general, the situation is not this simple.

\citet[][]{zirea13} recently showed that to explain the available
multi-wavelength  observations of Cas A, nonlinear DSA (with a total
efficiency $\gsim$ 25\%) of electrons, protons, and oxygen
ions, by both the forward and reverse shocks, must be invoked.
Particle injection in their kinetic models is an  adjustable
parameter and was varied independently for the forward and reverse
shocks to fit the multi-wavelength observations.
The  model results in a variety of
spectral shapes  -- hard spectra for oxygen
ions accelerated in the reverse shock and soft for proton spectra from the
forward shock with spectral indexes above 2, as inferred for Tycho.
Furthermore, soft CR spectra might be expected for
quasi-perpendicular shocks where the  angle between the shock normal
and the ambient magnetic field $\sim 90\degg$
\citep[e.g.,][]{SchureBell2013}.
Another uncertainty occurs for DSA in partially ionized plasmas where the shock compression is reduced by
the neutral return flux \citep[e.g.,][]{blasiea12,blasi13}.
We cite these cases to emphasize that the
interpretation of gamma-ray emission from DSA in young SNRs requires
careful modeling of complicated systems and is not yet fully developed.

\section{Discussion and Conclusions} \label{Discuss}
We have presented a comprehensive, \NL\ model of magnetic field amplification in shocks  undergoing
DSA.
The magnetic turbulence responsible for scattering particles is calculated \SCly\ from the free energy in
the anisotropic, position dependent, distributions of those
particles.
For the first time, we simultaneously include turbulence growth from the resonant CR streaming instability together with the
\nonres\ short-- and long--wavelength, CR--current--driven instabilities.
From the magnetic turbulence, we determine the particle diffusion coefficient with a set of assumptions that depend on the particle momentum.
Our plane-parallel, steady-state model includes shock modification and thermal particle injection  and, for the parameters assumed here, results in  efficient acceleration with
$\Qesc$ up to $\sim 50\%$ of the incoming energy flux and large downstream CR pressures.
We have explored the implications of this efficient acceleration with a limited parameter survey.

\newlistDE
Despite the approximations required,
we believe this is the most general description of NL DSA yet presented for the following reasons: \\
\noindent\listDE The \mc\ technique has a wide dynamic range and can follow particles from injection at thermal energies ($\sim 1$\,eV) to escape at PeV energies. This range is currently far greater than can be achieved with PIC simulations or hydrodynamic turbulence calculations.
Since no assumption of near-isotropy is required, both the injection of thermal particles and the escape of the highest energy CRs can be treated \SCly\ with a determination of the shock structure.
Modeling the feedback between thermal and PeV particles is essential in high Mach number shocks typical of young SNRs since CRs near $\pmax$ can contain a large fraction of the shock ram
kinetic energy;\\
\noindent\listDE  Our iterative model is the first to include the combined growth rates for resonant and non-resonant CR-driven
instabilities and to derive these from the highly anisotropic CR distributions in the \SCly\ determined shock precursor.
The instabilities produce highly amplified magnetic fluctuations which feed on one another making a consistent treatment essential;\\
\noindent\listDE  The iterative technique also provides a way to calculate  the scattering center speed $\Vscat(x)$, directly from energy conservation without assuming any particular form for the magnetic turbulence. This approach is different from all previous treatments of the effects of a finite scattering center speed on DSA. Since the CR current modifies the dispersion properties of the magnetic modes  and results in
fast non-resonant instabilities in the shock precursor, there is no reason to assume the self-generated turbulence in NL shocks is well described as \alf\ waves. We do not  assume this  and believe ours is the first attempt at a general determination of $\Vscat$ since the resonant \alf\ wave instability  was introduced for NL DSA
by \citet{McKVlk82}.\\

Other non-test-particle techniques that are currently being used to study NL DSA and MFA
with instabilities in addition to the resonant streaming instability are MHD calculations or plasma simulations,
either full-particle PIC or hybrid.

Three-dimensional MHD calculations have been performed by
\citet{BellEtal2013}
\citep[see also][and references therein]{bsr11,SB13} where the CRs are modeled kinetically and the CR current responsible
for the non-resonant hybrid (NRH)
(i.e., Bell) instability is included \SCly. Particular attention is given to the escaping CRs in this work and, based on estimates for the escaping CR current needed to produce enough NRH turbulence to confine high-energy CRs, the model predicts the maximum CR energy without scaling by the age or size of the accelerator \citep[see, for example,][for earlier work on determining $\pmax$ taking into account \NL\ effects]{bac07}. Due to the computational requirements of the 3D MHD simulation, however, the CR momentum range is restricted to a factor of order 10--100 and large shock speeds ($c/5$) are assumed.

As is well known,  plasma simulations have a great advantage over other techniques because, in principle, they provide a full, \SC\ calculation of the shock formation, particle injection and acceleration, and magnetic turbulence generation.
This comes with severe computational requirements imposed by directly following particles in a self-generated magnetic field where the relevant length and time scales are set by the mircoscopic plasma parameters.
For a hybrid simulation, the basic length scale is the ion inertial length, $c/\Opi$, and the basic time scale is the ion gyro-period
$t_g = m_pc/(eB_0)$.

Recently, \citet{CS2014}
\citep[see also][]{CS2013,CSeff2013}
have presented hybrid simulations of large, high Mach number, parallel shocks.
They see some evidence for the LWI
and suggest this may be from escaping CRs.
As we noted in our discussion of Figure~\ref{fig:Cap}, however, we see a very different MFA scaling from their equation~(2). While it is not clear what causes this, the basic assumptions and scales of the two simulations are extremely different.
For the time-dependent PIC simulations of \citet{CS2014}, for a background field $B_0=3$\,\muG, the largest dimension
in a two-dimensional box was $\Lmax = 4\xx{5} c/\Opi \simeq 5\xx{-6}$\,pc, the longest run time was $\Tmax = 2500 t_g \simeq 3\xx{-3}$\,yr, and the momentum range was less than three decades.
For the run with  $\Lmax = 4\xx{5} c/\Opi$, the transverse size
was $1000\,\Opi \simeq 10^{-8}$\,pc.
In comparison, the steady-state \mc\ shocks used to generate
Figure~\ref{fig:Cap} had $\Lfeb \simeq 0.11$\,pc, followed CR acceleration for the equivalent of 100's of years, and had a momentum range of $\sim 8$ decades.
Regardless of the computational limits of the plasma simulations, they contain critical plasma physics that is not modeled with
the \mc\ or MHD techniques making it important to
meaningfully match  the ``small-scale" plasma simulation results to the ``large-scale" \mc\ results to obtain  a consistent calculation over a dynamic range large enough in both space and time to model SNRs.

The results we have presented are all for highly efficient DSA
since our main goal is to study the effects of MFA from the
three CR-driven instabilities in such cases.
While global CR acceleration efficiencies on the order of $\simeq 10\%$ are generally assumed for SNRs, and efficiencies on this order are obtained by simpler parameterized NL models
\citep[e.g.,][]{ESPB2012}, local efficiencies could be far higher \citep[e.g.,][]{VBK2003}.
There is also indirect evidence from
multi-wavelength observations of some SNRs for high efficiencies
\citep[see, for example,][]{HRD2000,reynolds08,helderea12}.
We note that a direct {\it observation} of CR acceleration
efficiency $\simeq 25\%$ was obtained at the small, weak
quasi-parallel Earth bow shock \citep[][]{EMP90} and we see  no
fundamental reason, either from theory or SNR observations, that
restricts the intrinsic acceleration efficiency to values well below
what we model at large,  strong, SNR shocks, at least in some
circumstances.

The efficiency of DSA, at least  in terms of the downstream $\Pcr$, is potentially observable through the widths of Balmer lines from young SNRs in partially ionized material. This gives an estimate of the electron temperature and a model dependent estimate of the ion temperature and CR pressure can be obtained in association with
X-ray observations and a measurement of the shock velocity. We note that neutrals may play an important role in the energy balance of shocks with velocities below $\sim 3000$\,\kmps\  propagating in partially ionized
media \citep[see][for reviews]{helderea12,blasi13,bmrkv13}.
Here, we assumed fully ionized plasmas and did not attempt to model the effects of neutrals. To directly compare the pressure in the trapped
downstream CR particles with that derived from Balmer line
observations, one should account for the effects of neutrals and the
realistic geometry of the postshock flow in a \NL\ DSA model.

Of course the spectral shape is an additional  fundamental
prediction of DSA and concave spectra as  hard as we show may
present a problem in this regard, as mentioned above in our
discussion of $\Vscat$.
We caution that, as is generally the case, we make a diffusion approximation for CR
propagation.\footnote{While the \mc\ model doesn't rely on a diffusion equation to describe the evolution of the CR distribution particles are propagated with an assumed exponential pathlength distribution typical of ordinary diffusion.}
However, \citet[][]{bsr11} showed that  diffusive propagation may  break down for CRs in oblique shocks.
They found that higher order anisotropies, that appear in a
non-diffusion model, may result in harder spectra at quasi-parallel
shocks and softer spectra at quasi-perpendicular shocks.
It is important to point out that departures from standard diffusive propagation may be important in \NL\ shocks and such propagation may modify the spectral shape. We shall consider such departures in a seperate paper.

\acknowledgements
The authors thank the referee for useful comments. D.C.E acknowledges support from NASA grant NNX11AE03G. A.M.B. was partially supported by the Russian Academy of Sciences  OFN 15 Program. S.M.O. was partially supported by  the Russian Academy of Sciences  Presidium Program, RBRF grant for young 
scientists 14-02-31721.

\bibliographystyle{aa}
\bibliography{bibliogr}

\clearpage

% tttt
\begin{table}
\begin{center}
\caption{Model Parameters.$^\dagger$}
\label{tab:param}
\vskip6pt
\begin{tabular}{crrrrrrrrrrrrrr}
\tableline \tableline
\\
%
%%%
Model
& Inst.\tablenotemark{a}\tablenotetext{1}{The letter B stands for Bell's instability, L stands for the long-wavelength instability, and C indicates that cascading is included. All models include the resonant instability.}
&$u_0$
&$n_0$
&$\Lfeb$\tablenotemark{b}\tablenotetext{2}{The distance $\Lfeb$ is determined from a given number, $N$, of $\rgz$ such that
$\Lfeb \simeq 1.1\xx{-9} N
\left( \frac{u_0}{1000\,\mathrm{km/s}} \right)
\left( \frac{3 \mu\mathrm{G}}{B_0} \right)$\,pc. }
&$M_S:M_A$
&$\Rtot$:$\Rsub$
&$\PcrDS$\tablenotemark{c}\tablenotetext{3}{Percent of far upstream momentum flux, $\PhiPz$.}
&$\PwDS$\tablenotemark{c}
&$\Qesc$\tablenotemark{d}\tablenotetext{4}{Percent of far upstream energy flux, $\PhiEz$.}
%
%55 &$\EffDSA$\tablenotemark{d}
%
&$\BeffDS$
&$T_2$
\\
&
&
\kmps
&\pcc
&pc %%($\rgz$)
&
&
&\%
&\%
&\%
%55 &\%
&\muG
&$10^6$\,K
\\
\tableline
\aafourUM\tablenotemark{e}\tablenotetext{5}{Model UM is an unmodified shock where energy amd momentum are not conserved.}
&\dots
&5000
&0.3
&0.56 %%($1\xx{8}$)
&$430:420$
&$4.0:4.0$
&\dots
&\dots
&\dots
%55 &\dots
&3
&490\tablenotemark{f}\tablenotetext{6}{Upstream temperature, $T_0=10^4$\,K.}
\\
\aafour
&B, L
&5000
&0.3
&0.56 %%($1\xx{8}$)
&$430:420$
&$11.3:2.85$
&73
&9
&37
%55 &90
&540
&30\tablenotemark{f}
\\
\aafive
&B, L, C
&5000
&0.3
&0.56 %%($1\xx{8}$)
&$430:420$
&$13.3:2.52$
&80
&3.5
&47
%55 &95
&330
&20\tablenotemark{f}
\\
\tableline
\aathree
&B, C
&5000
&0.3
&0.56 %%($1\xx{8}$)
&$430:420$
&$12.5:2.37$
&79
&3.5
&42
%55 &94
&340
&20\tablenotemark{f}
\\
\aatwo
&B
&5000
&0.3
&0.56 %%($1\xx{8}$)
&$430:420$ &$10.2:2.84$ &69 &10 &35
%55 &87
&570
&35\tablenotemark{f}
\\
\aaeleven
&B, L
&2500
&0.3
&0.11 %%($4\xx{7}$)
&$210:210$
&$11.7:2.81$
&76
&6
&41
%55 &92
&220
&6\tablenotemark{f}
\\
\tableline
\aaseven
&B, L
&5000
&0.3
&0.11 %%($2\xx{7}$)
&$430:420$
&$11.0:2.83$
&72
&10
&37
%55 &89
&560
&30\tablenotemark{f}
\\
\aaone
&B, L
&10000
&0.3
&0.11 %%($1\xx{7}$)
&$850:840$
&$8.3:2.85$
&61
&12
&23
%55 &80
&1250
&220\tablenotemark{f}
\\
\aasixteen
&B, L
&20000
&0.3
&0.11 %%($5\xx{6}$)
&$1700:1670$
&$6.93:2.87$
&52
&14
&15
%55 &72
&2600
&1400\tablenotemark{f}
\\
\tableline
\aasix
&B, L
&5000
&1
&0.11 %%($2\xx{7}$)
&$430:760$
&$10.3:2.88$
&69
&11
&36
%55 &87
&1100
&35\tablenotemark{f}
\\
\aaeight
&B, L
&5000
&3
&0.11 %%($2\xx{7}$)
&$430:1300$
&$9.2:2.92$
&65
&11
&29
%55 &84
&1900
&45\tablenotemark{f}
\\
\aanine
&B, L
&5000
&10
&0.11 %%($2\xx{7}$)
&$430:2400$ &$8.65:3.01$ &62 &13 &25
%55 &81
&3700
&50\tablenotemark{f}
\\
\tableline
\aaten
&B, L
&5000
&30
&0.11 %%($2\xx{7}$)
&$430:4200$ &$8.58:3.10$ &60 &14 &23
%55 &80
&6500
&55\tablenotemark{f}
\\
\aafourteen
&B, L
&5000
&0.3
&0.34 %%($6\xx{7}$)
&$430:420$
&$11.0:2.83$
&72
&9
&39
%55 &90
&520
&30\tablenotemark{f}
\\
\bbthree
&B, L
&1000
&0.3
&0.11
&$8.5:84$
&$6.78:1.45$
&66
&1.5
&12
%55 &83
&41
&3.3\tablenotemark{g}\tablenotetext{7}{Upstream temperature, $T_0=10^6$\,K.}
\\
\tableline
\aaseventeen
&B, L
&30000
&0.3
&0.11 %%($3.3\xx{6}$)
&$2560:2500$ &$6.33:2.85$ &48 &13 &11
%55 &67
&3900
&3800\tablenotemark{f}
\\
\bbfour
&B,L,C
&1000
&0.3
&0.11
&$8.5:84$
&$6.72:1.33$
&66
&0.3
&19
%55 &83
&19
&3.5\tablenotemark{g}
\\
\bbeight
&B,L,C
&2500
&0.3
&0.11
&$21:210$
&$10.4:1.94$
&75
&1.0
&31
%55 &93
&90
&7.9\tablenotemark{g}
\\
\tableline
\bbsix
&B,L,C
&5000
&0.3
&0.11
&$43:420$
&$11.9:2.16$
&79
&3
&44
%55 &94
&315
&24\tablenotemark{g}
\\
\bbtwo
&B,L,C
&10000
&0.3
&0.11
&$85:840$
&$10.9:2.34$
&75
&5.3
&39
%55 &91
&820
&110\tablenotemark{g}
\\
\bbten
&B,L,C
&20000
&0.3
&0.11
&$170:1700$
&$9.15:2.43$
&68
&8
&28
%55 &86
&2000
&710\tablenotemark{g}
\\
\tableline
\bbtwelve
&B,L,C
&30000
&0.3
&0.11
&$260:2500$
&$8.0:2.47$
&63
&8.5
&23
%55 &82
&3100
&2200\tablenotemark{g}
\\
\bbeleven 
&B,L 
&30000 
&0.3 
&0.11 
&$260:2500$ 
&$6.33:2.85$ 
&48 
&13
&11
&3900
&3800\tablenotemark{g}
\\
\bbseven
&B,L 
&2500 
&0.3 
&0.11 
&$21:210$ 
&$10.5:2.13$ 
&75 
&5
&39
&200
&7.3\tablenotemark{g}
\\
\tableline
\bbfive
&B,L 
&5000 
&0.3 
&0.11 
&$43:420$ 
&$10.5:2.6$ 
&72 
&8.4
&37
&520
&29\tablenotemark{g}
\\
\bbone
&B,L 
&10000
&0.3 
&0.11 
&$85:840$ 
&$8.31:2.79$ 
&61
&12.4
&23
&1250
&215\tablenotemark{g}
\\
\bbnine
&B,L 
&20000
&0.3 
&0.11 
&$170:1700$ 
&$6.84:2.83$ 
&52
&13
&14
&2600
&1400\tablenotemark{g}
\\
\tableline
\end{tabular}
\end{center}
$^\dagger$For all models, $B_0=\Bkolm=3\xx{-6}$\,G and  we note that the derived results for all runs have a statistical uncertainty of $\lsim 5\%$.
\end{table}
%above tttt

\end{document}